\documentclass[aps,prd,reprint,groupedaddress]{revtex4-1}
\usepackage{amsmath,amsfonts,amssymb,amscd,amsthm,xspace}
\usepackage{graphicx}
\usepackage{epstopdf}
\usepackage{enumerate}
\usepackage{wrapfig}
\usepackage{color}
\usepackage{longtable}
\usepackage{float}
\usepackage{hyperref}

\usepackage{times}

\begin{document}

%\preprint{}

\title{Relativistic dust accretion of charged particles in Kerr--Newman spacetime}


\author{Kris Schroven}
\email[]{kris.schroven@zarm.uni-bremen.de}
\author{Eva Hackmann}
\email[]{eva.hackmann@zarm.uni-bremen.de}
\author{Claus L\"ammerzahl}
\email[]{claus.laemmerzahl@zarm.uni-bremen.de}
%\thanks{}
\affiliation{University of Bremen, Center of Applied Space Technology and Microgravity (ZARM), 28359 Bremen}


\date{\today}

\begin{abstract}
We describe a new analytical model for the accretion of particles from a rotating and charged spherical shell of dilute collisionless plasma onto a rotating and charged black hole. By assuming a continuous injection of particles at the spherical shell and by treating the black hole and a featureless accretion disc located in the equatorial plane as passive sinks of particles we build a stationary accretion model. This may then serve as a toy model for plasma feeding an accretion disc around a charged and rotating black hole. Therefore, our new model is a direct generalization of the analytical accretion model introduced by E. Tejeda, P. A. Taylor, and J. C. Miller (2013). 

We use our generalized model to analyze the influence of a net charge of the black hole, which will in general be very small, on the accretion of plasma. Within the assumptions of our model we demonstrate that already a vanishingly small charge of the black hole may in general still have a non-negligible effect on the motion of the plasma, as long as the electromagnetic field of the plasma is still negligible. Furthermore, we argue that the inner and outer edges of the forming accretion disc strongly depend on the charge of the accreted plasma. The resulting possible configurations of accretion discs are analyzed in detail.
\end{abstract}

%\keywords{}

\maketitle


\section{Introduction}
    Accretion onto black holes (BHs) is a fundamental astrophysical process as it gives rise to a large range of astrophysical phenomenon like active galactic nuclei (AGN), X-ray binaries and gamma ray bursts \cite{frank,yuan}. 
  
    To describe the full picture of BH accretion one has to consider general relativistic magnetohydrodynamics, including turbulences, radiation processes, nuclear burning and more. The simulation of accretion processes therefore contains a number of challenging issues. It requires a large range of scales because some important effects, like the magnetorotational instability, only occur on very small scales, while for the interpretation of observational phenomena typically what happens on very large scales is of interest. The computational expense rises even more and by a large factor, if the number of dimensions which have to be taken into account increases, say, from one dimension (spherical model) to two (axis symmetric model) or to three. Therefore, it is necessary to reduce the computational costs by different methods and assumptions. The relevant number of dimensions can be reduced (eg. assuming axis symmetry), or the range of scales, which have to be taken into account (eg. shearing--box simulations). Negligence of certain aspects of the accretion process, like conduction, viscosity, or kinetic effects, simplifies the system of equations and leads to a reduction of the computational costs as well. 

    To understand the general physical processes, analytical models of the accretion process play a very important role. Besides serving as initial conditions or test beds, analytical models are indispensable to understanding the resulting observational features of the accretion process which have to be taken into account for numerical simulations. An early work discussing observational features is by \citet{michel}, who generalized in his analytical model the spherical accretion model of \citet{bondi} to the general relativistic case and gave the first estimates for the realized temperatures and luminosities in the accretion of a polytropic gas. Furthermore by assuming the polytropic gas to be a plasma, estimations for the strength of the arising electrostatic field were discussed.
    
    This simplest case of spherical accretion, however, was found to have a low efficiency in converting potential energy to radiation  \citep{shapiro}, which is why the rotation of accreted matter was invoked in accretion theories.
    
    Rotating inflows suggest the existence of accretion discs, introduced among others by \citet{prendergast}. Accretion discs and the processes within are discussed extensively in literature, by introducing different (analytical) models to describe them, such as thin discs, slim discs, Polish Doughnuts, advection-dominated accretion flows (ADAFs), and more (see \citep{abramowicz}, and citations within). These accretion disc models significantly advanced our understanding of the accretion process, and can therefore be used to enhance numerical simulations. They are further used to understand specific observational results, such as the truncated disc model, built by a truncated thin disc adjoined with an inner ADAF-like flow \citep*{done}. 
    
    Cosmic matter mainly exists in the form of plasma. It serves as the main ingredient of stars, interstellar nebulae, solar wind, jets, and AGN \citep{buchplasma,buchchen}. Therefore, it is reasonable to assume that the matter accreted by a massive central object is some form of plasma. The broad range of temperatures and densities (from $<0.01$ to $>10^5\;\rm{cm}^{-3}$ in ultra compact HII regions \citep{himcox,uchii}) in which plasma may occur can be taken into account using different plasma models. This includes hot and cold plasma, or plasma with and without taking into account particle collisions (collisional or collisionless plasma). The different descriptions range from plasma described as a fluid over a kinetic theory of plasma to a description of it as a collection of individual particle motions \citep{buchchen}.

    Plasma accretion is also one of the reasons why in realistic astrophysical models the net charge of the accreting BH is expected to be very small. Selected accretion of oppositely charged particles will reduce the net charge to a very small value within a short time scale \citep{charge3}. In the case of stellar BHs this will even happen in vacuum due to pair production \citep{charge1,charge2}. In these scenarios the influence of the net charge on the spacetime geometry is therefore vanishingly small. However, we will show that the remaining charge can still be strong enough to have a noticeable influence on the motion of charged test particles. Note that there are also accretion scenarios which may create BHs with a net charge big enough to have influence on the spacetime geometry \citep{bigc1,bigc2,bigc3}.
    
    Here, we will discuss the relativistic accretion of plasma by a rotating BH with a (very small) net charge. We restrict to the accretion from a rotating cloud of dust, thereby generalizing the analytical model introduced by \citet*{mendoza}, \citet*{accss} and \citet*{acck} in a Newtonian approach and for the Schwarzschild and Kerr spacetimes, respectively. In these references it was shown that this model is well suited to explore relativistic effects, such as frame dragging, on the accretion process and may be used in numerical simulations for collapsarlike setups to reduce computational costs. For the model of the plasma we restrict to a collisionless dilute plasma, i.e. in the form of a collection of individual charged particles. Our analytical model will help to understand the influence of specific angular momentum and net charge of the BH on the accretion process of charged dust. It might also serve as a toy model for the infall of plasma feeding an accretion disc around a charged and rotating BH. 
  
    In General Relativity rotating and charged BHs are described by the Kerr--Newman \citep{kerrnew} metric, which is a generalization of the Kerr metric. Besides an electric charge it also allows the consideration of a magnetic net charge. However the existence of magnetic monopoles, in general, was never proven and we will not consider the magnetic charge further here.
    
    Our analytical model necessarily simplifies the complex physical processes involved in the accretion. In particular, we assume stationarity, axial symmetry, and the absence of particle interaction. As a result, pressure gradients within the accreted plasma are neglected, as well as self gravity. The charged particles are also assumed to only interact with the gravitational and the electromagnetic field of the BH, and we neglect the electromagnetic field produced by the plasma particles itself. This will restrict the particle density of the accreted cloud, especially in case of a central BH with a very small net charge. Within this relativistic model the trajectories of the individual charged particles which form the plasma can then be analytically described, see \citep{acck,hackmann}. This allows to clearly analyze effects which are purely relativistic or caused by the interaction with the electromagnetic field of the BH. 
  
    The paper is organized as follows. First an introduction of the Kerr--Newman spacetime and the equations of motion for charged test particles are given in Sec. \ref{ecom}. Then we explain the relativistic analytical model of the accretion used in this paper, including restrictions to the initial conditions in Sec. \ref{setup}. In Sec. \ref{features} we discuss the accretion flow, with details on the velocity field in locally nonrotating reference frames (LNRFs), a description of the streamlines in terms of Jacobi elliptic functions, a derivation of the inner most stable orbit (ISCO) in Kerr--Newman spacetime which corresponds to the inner edge of an accretion disc in our model, and the calculation of the density field formed by the accreted matter. In Sec. \ref{results} the results are summarized and discussed. Finally, we conclude in Sec. \ref{conclusion}.
    
\section{Equations of motion in Kerr--Newman spacetime \label{ecom}}
    The Kerr--Newman spacetime is a stationary and axially symmetric solution of the Einstein--Maxwell equation, which describes a charged rotating BH \citep{kerrnew}. It allows us to consider both electric and magnetic net charges, however, we will not consider a magnetic charge of the BH here. In the Boyer--Lindquist system of coordinates $\left(t, r,\phi,\theta\right)$ the Kerr--Newman metric takes the form
        \begin{align}
	  ds^2=&\frac{\rho^2}{\Delta}dr^2+\rho^2d\theta^2+\frac{\sin^2\left(\theta\right)}{\rho^2}\left[\left( r^2+a^2\right)d\phi -a\;c\;dt\right]^2 \nonumber \\
             &-\frac{\Delta}{\rho^2}\left[a \sin^2\left(\theta\right)d \phi-c\;dt\right]^2,
        \end{align}
        where
        \begin{align}
        \rho^2(r,\theta)=r^2+a^2 \cos^2\left(\theta\right),\\
        \Delta(r)=r^2-2 M r+a^2 + Q^2 + P^2.
        \label{delta}
        \end{align}
   Here the parameters $M$, $a$, and $Q$ are related to the angular momentum $J$, the mass $m$, and the electric charge $q$ of the BH by
     \begin{align}
       \label{parize1}
       a=&\frac{J}{mc},\\
       \label{parize2}
       M=&\frac{Gm}{c^2},\\
       Q^2=&\frac{q^2 G}{4\pi\varepsilon_0 c^4 },
       \label{parize3}
      \end{align}
   where $G$ is Newton's gravitational constant, $c$ is the speed of light and $\varepsilon_0$ is the electric constant. The parameter $P$ corresponds to the magnetic monopole.

   The Kerr--Newman spacetime has two horizons $r_\pm$, which are located at the coordinate singularities $\Delta(r)=0$, i.e. $r_{\pm}=M \pm \sqrt{M^2 - a^2 -  Q^2 - P^2}$. The curvature singularity is given by $\rho(r,\theta)=0$, i.e. at simultaneously $r=0$ and $\theta=\frac{\pi}{2}$ which corresponds to a ring singularity. In the following we will only consider the region of the spacetime outside the event horizon, $r>r_+$.
   
   The electromagnetic potential is
    \begin{align}
    A= A_\nu dx^\nu=&\frac{c^2}{\sqrt{4\pi\varepsilon_0 G}}\left\{\frac{Q r}{\rho^2}\left(dt-a \sin^2\left( \theta\right) d\phi \right) \right.\nonumber\\
                    &\left. +\frac{1}{\rho^2} P \cos\left( \theta\right)\left(a dt-(r^2+a^2)d\phi \right)\right\}\\
                   =&\frac{c^2}{\sqrt{4\pi\varepsilon_0 G}} \bar{A}_\nu dx^\nu .
    \end{align}

     We now consider the motion of test particles with a mass $\mu$ which is very small compared to $m$ and a specific electric charge parameter $\hat{e}=e/\mu$ which is related to the charge $\epsilon$ of the particle by
     \begin{align}
     \label{repe} e=&\frac{\epsilon}{\sqrt{4\pi\varepsilon_0 G}}.
     \end{align}
     The Hamilton--Jacobi equation for such a charged particle in Kerr--Newman spacetime is separable and leads to the equations of motion and four separation constants. Equivalently, one can also derive the equations of motion directly. We first note that the Hamiltonian of a charged test particle does not depend on $\phi$, $t$, or proper time $\tau$, which can be used to obtain three constants of motion directly. We find the four-velocity modulus, the specific energy $E$, and the specific angular momentum in $z$ direction $l$ as
    \begin{align}
    \label{com0}
    u^\mu u_\mu&=-c^2,\\
    \label{com1}
     E = \frac{\textbf{E}}{\mu c^2}  & = -g_{00}\dot t -g_{0\phi}\frac{\dot\phi}{c}+\hat{e} \bar{A}_t,\\
    \label{com2}
     l =\frac{L}{\mu c}   & = g_{\phi 0}\dot t +g_{\phi\phi}\frac{\dot\phi}{c}-\hat{e} \bar{A}_\phi,
    %\hat K = \frac{K}{\mu^2}& = \rho^2\dot\theta^2+\hat l^2 \cot^2\left(\theta\right)-\left(\hat E^2-1\right)a^2\cos^2\left(\theta\right)+\left( a\hat E-\hat l \right)^2.
    %\label{com3}
    \end{align}
     where zeroth component of the four vector is is defined as $x^0=c\,t$. Here the dot denotes a differentiation with respect to proper time $\tau$. We may now solve Eq. \eqref{com1} and \eqref{com2} for $\dot \phi$ and $\dot t$ and find the first two equations of motions. If we introduce the Mino time $\lambda$ via $d\lambda= \rho^{-2}d\tau$ \cite{mino} they take the form
     \begin{align}
     \frac{1}{c}\frac{d \phi}{d\bar\lambda}                   &= \frac{\bar a \mathcal{R}\left(\bar r\right)}{\bar \Delta\left(\bar r\right)}-\frac{\mathcal{T}\left(\theta\right)}{\sin^2\left(\theta\right)},\\
     \frac{d \bar t}{d \bar\lambda}                     &= \frac{\left(\bar r^2+\bar a^2\right)\mathcal{R}\left( \bar r\right)}{\bar \Delta\left(\bar r\right)}-\bar a\mathcal{T}\left(\theta\right),
     \end{align}
     where
     \begin{align}
     \label{curvyt}
     \mathcal{T}(\theta)                      &= \bar a E\sin^2{\theta} - \bar l+\hat e \bar P \cos{\theta},\\
     \label{curvyr}
     \mathcal{R}(\bar r)                      &= (\bar r^2+\bar a^2)E-\bar a \bar l-\hat e \bar Q \bar r.
     \end{align}
     Here we eliminated $M$ from the equations by using the transformation $x=\bar x M$ for $x=r,t,a,l,Q,P,d/d\lambda$. By inserting the equations for $\phi$ and $t$ into Eq. \eqref{com0} and by using again the Mino time, \eqref{com0} becomes separable for $r$ and $\theta$ and we find
     \begin{align}
    \label{introK}
    \bar K=&\left(\frac{1}{c}\frac{d \theta}{d \bar\lambda}\right)^2+  \bar a^2\cos^2{\theta}+\frac{\mathcal{T}^2(\theta)}{\sin^2{\theta}} \nonumber\\
    =& \frac{1}{\bar \Delta (\bar r)}\left(\mathcal{R}^2(\bar r)-\left(\frac{1}{c}\frac{d \bar r}{d \bar\lambda}\right)^2\right)-\bar r^2\,.
     \end{align}
     The separation constant $K=\bar K M^2$ is the fourth constant of motion. It is connected to the Carter constant $C$, which was found by Carter in 1968, by $C=K-(aE-l)^2$. Summarized we find
     \begin{align}
     \label{dqth}
     \frac{1}{c^2}\left(\frac{d \theta}{d \bar\lambda}\right)^2 &= \bar K-\bar a^2\cos^2{\theta}-\frac{\mathcal{T}^2(\theta)}{\sin^2{\theta}}=\Theta\left(\theta\right),\\
     \label{dqr}
     \frac{1}{c^2}\left(\frac{d \bar r}{d \bar\lambda}\right)^2 &= \mathcal{R}^2(\bar r)-(\bar r^2+\bar K)\bar \Delta (\bar r)=\textbf{R}(\bar r),\\
     \label{dqph}
     \frac{1}{c}\frac{d \phi}{d\bar\lambda}                   &= \frac{\bar a \mathcal{R}\left(\bar r\right)}{\bar \Delta\left(\bar r\right)}-\frac{\mathcal{T}\left(\theta\right)}{\sin^2\left(\theta\right)},\\
     \label{dqt}
     \frac{d \bar t}{d \bar\lambda}                     &= \frac{\left(\bar r^2+\bar a^2\right)\mathcal{R}\left( \bar r\right)}{\bar \Delta\left(\bar r\right)}-\bar a\mathcal{T}\left(\theta\right).
     \end{align}

     In the following we will use $c=1$ and skip the bars for all parameters and variables, if not explicitly noted otherwise.

\section{The Model of accretion  \label{setup}}
\begin{figure}
	\centering
	\includegraphics{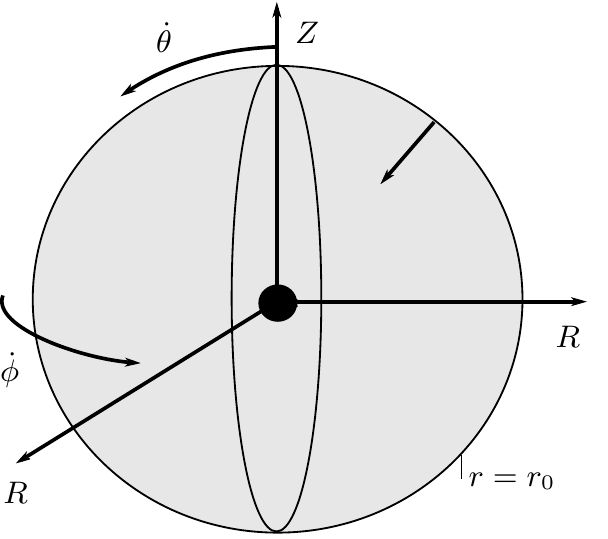}
        \caption{Sketch of the accretion model. Here $R=\sqrt{r^2 +a^2}\sin(\theta)$ and $Z=r \cos(\theta)$.}
       \label{skizze}
    \end{figure}

   In the following the accretion model will be introduced in more detail. It basically consists of three parts: (i) a rotating and charged BH, which solely determines the gravitational and electromagnetic field, (ii) a featureless accretion disc, lying in the equatorial plane, and (iii) a rotating and charged spherical shell of particles located at a certain radius $r_0$, which is continuously fed with new particles. A sketch of the model is given in Fig. \ref{skizze}.

  \paragraph*{The cloud of particles}
    The particle cloud is assumed to form a plasma (if charged) and to be sufficiently dilute, such that particle collisions can be neglected, and the electromagnetic field of the particle cloud is negligible compared to the field of the BH. This leads to a ballistic accretion flow and a collisionless plasma. Furthermore, we assume that the electromagnetic and gravitational field formed by the plasma can be neglected compared to the field of the central BH. In this case the trajectory of each individual particle in the cloud, charged or uncharged, follow a path of test particles in the given spacetime as described by the equations of motion \eqref{dqth}--\eqref{dqt}. 
          
    The particles of the dust cloud are assumed to be continuously injected at $r_0$, where they have a constant $r$-, $\phi$- and $\theta$-velocity $\left( \dot r_0, \dot\phi_0, \dot\theta_0\right)$, and then start falling onto the BH and either hit the accretion disc or reach the event horizon. 
    
    As mentioned above we assume an accretion disc in the equatorial plane, which only makes sense if the spacetime exhibits a reflection symmetry with respect to the equatorial plane, defined by $\theta=\pi/2$. We discuss this in more detail below and just postulate this here. The initial conditions $\dot r_0, \dot\phi_0$, and $\dot\theta_0$ are chosen such that they reproduce this spacetime symmetry. Due to these symmetric initial conditions, particles starting at $r_0$ from the upper half plane will collide with their corresponding particle starting from the lower half plane precisely at the equatorial plane. By choosing the three initial conditions $\left( \dot r_0, \dot\phi_0, \dot\theta_0\right)$ the three constants of motion $E,L$ and $C$ are completely determined and can be calculated using Eqs. \eqref{com0}--\eqref{com2} and \eqref{dqth}. 
   
    It is required that there are no turning points in the streamlines, described by the $\theta$- and $r$-motion of the particles, before they reach the $\theta=\pi/2$ plane. Furthermore, the mapping 
    \begin{equation}
     	\left.\left(\frac{\partial \theta}{\partial \theta_0}\right)\right|_{r=const}\ge0
     	\label{thth0c}
    \end{equation}
    should hold. Otherwise, streamlines of particles with the same charge may interSec. In this case, Eq. \eqref{density}, which describes the arising density field, diverges at the points of streamline intersection. It is not an easy task to check, if this condition holds for given initial conditions. However, during calculation of the streamlines for various different initial conditions, we found that the main cause for intersecting streamlines are turning points in the $r$-motion for $\theta<\pi/2$.
    This can be checked rather easily for given initial conditions. In general, to prevent turning points the specific angular momentum $l$ and the charge product $eQ$ have to be chosen sufficiently small.
 
  \paragraph*{The black hole and the accretion disc}
    We assume that the central BH is described by the Kerr--Newman metric as introduced in Sec.~\ref{ecom}, neglecting, however, the case of a magnetic monopole. In this case the postulated reflection symmetry with respect to the $\theta=\pi/2$ plane is realized, and the equations of motion are simplified. (For $P\neq0$ the situation looks different. Since the symmetry with respect to the equatorial plane is broken in that case, there is no reason to assume that the accretion disc is located at $\theta=\pi/2$.)
    
    The choice of constant initial conditions for the particle cloud results in a constant accretion rate $\dot M$, which can be calculated by 
    \begin{equation} \dot M=-\int{\int{ n_0 \rho^2(r_0)\dot r_0\sin(\theta_0)d\theta_0 d\phi_0}} \end{equation}
    for a particle density $n_0$ at $r_0$. However, we assume sufficiently small time scales such that the mass change of the BH and of the accretion disc can be neglected in our model. The BH as well as the accretion disc then act as passive sinks for particles and energy, and a stationary accretion model is built. Within this stationary model we can also deduce specific features of the for now featureless accretion disc.   
   
    When discussing the case of charged particles or a plasma we will restrict to very small values of the charge $Q$ of the BH and a product $eQ$ of the order of $10^0$. This restriction results from the following considerations. On the one hand, it can be expected that BHs with bigger net charges are quite unlikely, see for example \citet{charge1,charge2, charge3}. On the other hand, we assume the plasma to consist of protons and electrons. By going back to the notation used in Sec. \ref{ecom}, the dimensionless charge $\hat{e}$ of both electrons and protons can be calculated, using Eq. \eqref{repe}, 
    \begin{equation}\hat{e}_{\text{electron,proton}} =\frac{1}{\sqrt{4 \pi \varepsilon_0G}}\left(\frac{\epsilon}{\mu}\right)_{\text{electron,proton}}.\end{equation} 
    This leads to
    \[\hat{e}_{\text{electron}}\approx -2.042~ 10^{21}\text{ and }\hat{e}_{\text{proton}}\approx 1.112~ 10^{18}~.\]
 
    Considering a value of $Q\approx 1$, all terms in the constants and equations of motion can be neglected, which are small compared to $\hat e$. When we assume sufficiently small initial conditions for the $\phi$ and $r$ motion, so that they are small compared to $\hat{e}$, the constants of motion reduce to
  
    \begin{align}
     \mathcal{E}&=\frac{E}{\hat{e}}\approx A_t=\frac{Qr_0}{\rho_0^2},\\
     \mathcal{L}          &=\frac{l}{\hat{e}}\approx-A_\phi\approx a\mathcal{E}\sin^2{\theta_0},\\
     \mathcal{K}&=\frac{K}{\hat{e}^2}\stackrel{\text{Eq.}\eqref{introK}}{\approx} 0.
    \end{align}
 
    With them we can derive approximate expressions for the equations of motion,
       
    \begin{align}
     \label{tm}
     \left(\frac{1}{\hat{e}}\right)^2\left(\frac{d \theta}{d \lambda}\right)^2 &\approx -\frac{T^2(\theta)}{\sin^2{\theta}},\\
     \label{rm}
     \left(\frac{1}{\hat{e}}\right)^2\left(\frac{d \bar r}{d \lambda}\right)^2 &\approx R^2(r),\\
     \left(\frac{1}{\hat{e}}\right)^2\frac{d \phi}{d\lambda}                     &\approx \frac{ a R(r)}{\Delta(r)}-\frac{T(\theta)}{\sin^2{\theta}},
     \end{align}
     where
     \begin{align}
     \label{rfun}
      R(r)    =\frac{\mathcal{R}(r)}{\hat{e}}&\approx ( r^2+a^2)\mathcal{E}-a \mathcal{L}- Q r,\\
      T(\theta)  =\frac{\mathcal{T}}{\hat{e}}&\approx a\mathcal{E}\left(\sin^2{\theta} -\sin^2{\theta_0}\right).
    \end{align}       
    Eqn. \eqref{tm} can only be true for $T(\theta)=0$. This leads to a particle motion with a constant $\theta$ value, and thus to a radial infall, which is why this case is not of further interest in this paper. As a result, we will only consider very small values of $Q$ of the order of $10^{-18}$--$10^{-21}$, and we can neglect terms in Eqs.~\eqref{com0}--\eqref{curvyr}, which contain $Q$ but not $e$. 
    
    Note that these very small values of $Q$ still correspond to a comparibly large total net charge $q$ of the BH. According to Eq. \ref{parize3}, the total net charge per elementary charge $\epsilon$ is given by $\left|q\right|/\epsilon\approx 10^{21}$--$5\cdot10^{17}\,m/M_\odot$, where $m/M_\odot$ is the BH mass per solar mass. Hence, within our model, the accretion of electrons or protons will not significantly change the value of $Q$.
      
    Since we consider only protons and electrons as accreted particles, the value of the particle's charge $e$ is given by the elementary charge. Fixing the BH charge $Q$ therefore fixes the value of $eQ$, while on the other hand different values of $eQ$ correspond to different charges of the central BH. The sign of $eQ$ determines whether the particles and the BH have the same ($eQ>0$) or an opposite ($eQ<0$) charge.

\section{Features of the accretion process}
  \label{features}
  The description of the accretion process within our model is based on the analytical solutions of the streamlines and the velocity field, and on the numerical calculation of the density field of the accretion flow. In this section we introduce and discuss the equations covering this accretion process, based on the treatment presented in \citet{acck}. Furthermore we discuss the innermost stable orbit in Kerr--Newman spacetime, since it determines the inner edge of the accretion disc in our model.
  
  \subsection{The velocity field}
  \label{velfield}
    The components of the four-velocity $u^\mu=dx^\mu/d\tau$ are given by the equations of motion \eqref{dqth}--\eqref{dqt},
    \begin{align}
	u^r      &= \frac{\sqrt{\textbf{R}}}{\rho^2},\\
	u^\theta &= \frac{\sqrt{\Theta}}{\rho^2},\\
        u^\phi   &= \frac{a \mathcal{R}}{\rho^2 \Delta}-\frac{\mathcal{T}}{\rho^2\sin^2\theta},\\
	u^t      &= \frac{\left(r^2+ a^2\right)\mathcal{R}}{\rho^2 \Delta}- \frac{a\mathcal{T}}{\rho^2}.
    \end{align}  
    However, in order to obtain a local description of the velocity field, we will express it in a set of locally nonrotating frames. This set of reference frames was introduced by \citet*{lnrf}. It measures the velocity field seen by locally nonrotating observers, whose world lines are constant in $r$ and $\theta$, but change in $\phi$ with $\phi={\rm const.} +\omega t$ and $\omega=-\frac{g_{\phi t}}{g_{\phi \phi}}$. This means the observers are so to say ``frame-dragged''. The observers' orthonormal tetrads then locally constitute a set of Minkowskian coordinates.
    
    The components of the three velocity $(\frac{dr'}{dt},\frac{d\theta'}{dt},\frac{d\phi'}{dt})$ in the LNRFs are given by
    \begin{align}
      \label{vf1}
     \frac{dr'}{dt} = v^{ r'}     & = \frac{\sqrt{\textbf{R}/\Delta}}{\rho \gamma},\\
     \frac{d\theta'}{dt} = v^{\theta'} & = \frac{\sqrt{\Theta}}{\rho \gamma},\\ \label{vf3}
     \frac{d\phi'}{dt} =  v^{\phi'}   & = \frac{\rho\left(l-eQra\sin^2{\theta}\right)}{\gamma\sqrt{\left(r^2+a^2\right)^2-a^2\Delta\sin^2\theta}\sin{\theta}}
    \end{align}
    and
    \begin{equation}
    \gamma=\sqrt{1+{v^{ r'}}^2\gamma^2+{v^{ \theta'}}^2\gamma^2+{v^{ \phi'}}^2\gamma^2},  
    \label{vf4}
    \end{equation}
    where the magnetic monopole $P$ is already set to zero. Here $\gamma$ is the Lorentz factor between the LNRFs and the passing test particle. The expressions \eqref{vf1}--\eqref{vf3} for the velocity field contain the variables $(r,\theta)$ as well as the constants of motion $E$, $l$ and $K$, which depend on the initial values $r_0$, $\theta_0$, $\dot\theta_0$, $\dot r_0$, and $\dot\phi_0$ of the test particle. Therefore, to calculate the components of the velocity field we need to compute the variables $r$ and $\theta$ as functions of the initial conditions. These functional relations are provided in terms of streamlines.
    
  \subsection{Streamlines}
  \label{streamlines}
    Within our model the particles from the rotating shell will follow the motion of charged test particles in Kerr--Newman spacetime. Therefore, the streamlines of the accretion flow can be described by the solutions to the equations of motions  \eqref{dqth}--\eqref{dqt} in Kerr--Newman spacetime.

    As explained in Sec. \ref{setup}, our model has an axial symmetry to the $z$ axis. Therefore it is sufficient to consider the projection on the $(r,\theta)$ plane to fully discuss the streamlines of the particle motion. Furthermore, due to the reflection symmetry to the equatorial plane in our model, particles starting from the northern and the southern hemisphere will collide at $\theta=\pi/2$ and be absorbed by the accretion disc in the equatorial plane, which acts as a passive sink for particles. Therefore, we can further restrict our calculations to the upper half plane ($\theta<\pi/2$) of the $(r,\theta)$ plane.
    
    The equations of motion \eqref{dqth} and \eqref{dqr} can be solved by elliptic functions and integrals. A comprehensive discussion of the solutions of the Kerr--Newman equations of motions using Weierstrass elliptic functions was done by \cite{hackmann}. Here we use Jacobian elliptic functions to obtain the solution $r(\theta)$ for the streamlines in the $(r,\theta)$ plane. We will only write down the result at this point and refer to Appendix \ref{app1} for the derivation and more detailed explanations.

   The solution for $r(\theta)$ reads
    \begin{align}
      \label{rtheta}
      r(\theta) =& \frac{r_b(r_d-r_a)-r_d(r_b-r_a)\,\text{cn}\left(\xi, k_r\right)^2}{r_d-r_a-(r_b-r_a)\,\text{cn}\left(\xi, k_r\right)^2}
      \end{align}  
      with
     \begin{align}     
      \xi=& \frac{1}{2}\sqrt{(E^2-1)(r_a-r_c)(r_d-r_b)} \, \nonumber\\
          &\times \left[\Phi(r_0)+\Psi(\theta_0)-\Psi(\theta)\right]\,,
      \label{xi}
     \end{align}
     and
     \begin{align} 
      \Phi(r) & = 2\frac{\text{cn}^{-1}\left(\sqrt{\frac{(r_d-r_a)(r_b-r)}{(r_b-r_a)(r_d-r)}}, k_r\right)}{\sqrt{(E^2-1)(r_a-r_c)(r_d-r_b)}}\,,\\
      \Psi(\theta) & = \frac{\cos\theta_a\,\text{cn}^{-1}\left(\frac{\cos\theta}{\cos\theta_a}, k_{\theta}\right)}{\sqrt{C+(E^2-1)a^2\cos\theta_a^4}}.
    \end{align}
    Here $k_r$, $k_{\theta}$ are the moduli of the elliptic integrals given by
    \begin{align}
	k_r^2 & =\frac{(r_b-r_a)(r_d-r_c)}{(r_d-r_b)(r_c-r_a)}\,,\\
	k_{\theta}^2 & = \frac{a^2(E^2-1)\cos^4\theta_a}{C+a^2(E^2-1)\cos^4\theta_a}\,,
	\label{rmod}
    \end{align}
    $r_{a,b,c,d}$ are the four real or complex roots of $\textbf{R}(r)$, and $\theta_a$ is discussed below. The roots of $\textbf{R}(r)$ mark the turning points of the radial motion, since the motion can only take place where $\textbf{R}(r)$ is positive (see Eq. \eqref{dqr}). The roots are sorted differently, depending on between which roots of $\textbf{R}(r)$ the radial motion oscillates. We use the labeling of the roots introduced by \citet{acck}, which we shortly review here. 
    
    If all roots are real, two situations can happen: In the first case, the $r$-motion is bound between two non-negative roots of $R(r)$, called $r_a$ and $r_b$, for $r_a<r_b$. In the second case, the $r$-motion has a lower bound, $r_a$, but is unbounded above and $r_b$ is the root with the smallest value. In both cases the remaining roots are called $r_c,r_d$, with $\left|r_c\right|<\left|r_d\right|$. If two roots are real, and two roots form a complex conjugate pair, the real roots are called $r_a,r_d$, with $\left|r_a\right|<\left|r_d\right|$, and the complex roots are called $r_b,r_c$. If all roots are complex, one complex conjugate pair is called $r_a,r_d$ and the other one is called $r_b,r_c$.
    
    The root $\theta_a\in \left[0,\pi/2\right]$ of $\Theta\left(\theta\right)$ lies closest to the equatorial plane. Since the roots determine the turning points of the $\theta$-motion, $\theta_a$ sets the lower limit of the $\theta$-motion. In case of setting $\dot\theta_0$ to zero, $\theta_0$ and $\theta_a$ coincide.
    
    The form of the expression \eqref{rtheta} for the streamlines $r(\theta)$ does not differ from the one given in \citet{acck}. However, the position of the roots $r_{a..d}$ is influenced by the electric charge of the particles and the BH. Since the magnetic monopole is set to zero, the equation of motion for $\theta$ reduces to the one in Kerr spacetime and we recover the result for the $\theta$-motion as given in \citet{acck}, see Eq. \eqref{thetasol}.

    Please note that the constants of motions appearing in Eqs. \eqref{rtheta}--\eqref{rmod} are calculated by using 
    Eqs. \eqref{com1},\eqref{com2} and \eqref{introK}, for the initial values $r_0$, $\theta_0$, $\dot r_0$, $\dot\phi_0$, and $\dot\theta_0$. The value of $\dot t$ in these equations is determined by the condition in Eq. \eqref{com0}. As a consequence, the constants of motion are different for every streamline starting at $r_0$ with a different angle $\theta_0$.

  \subsection{The density field}
    \label{secdensity}
    \label{densityfield}
    To calculate the density field $n(r,\theta)$, we use the continuity equation
    \begin{equation}
	\left(nu^\mu\right)_{;\mu}=0.
    \end{equation}
    The semicolon denotes covariant differentiation. By using the Gauss theorem the continuity equation can be written as follows, 
    \begin{equation}
	\int_{\partial \mathcal{V}} n u^\mu N_\mu \sqrt{\left|h\right|} d^3 x=0.
    \end{equation}
    Here $N_\mu$ is a unit vector normal to the hypersurface $\partial \mathcal{V}$ delimiting the integration volume and $h$ is the induced metric's determinant on this hypersurface. By choosing the  infinitesimal integration volume 
    wisely, such that the spatial projection of $\partial \mathcal{V}$ is determined by neighboring streamlines and two area elements $\left. dx^2 \right|_{r_0}$, $\left.dx^2\right|_{r={\rm const.}}$, which are connected by the neighboring 
    streamlines, the following final equation can be deduced for the density field \citep{acck},
    \begin{equation}
	n=\frac{n_0u^r_0\rho_0^2\sin\theta_0}{u^r\rho^2\sin\theta}\left.\left(\frac{\partial\theta}{\partial\theta_0}\right)^{-1}\right|_{r={\rm const}},
	\label{density}
    \end{equation}
    where $n_0$, $u_0^r$, and $\rho_0$ are the values of $n$, $u^r$ and $\rho$ at $r=r_0$ and $\theta=\theta_0$. For the derivation of the equation above it was used that, by construction, particles will only flow through the area elements $\left. dx^2 \right|_{r_0}$ and $\left.dx^2\right|_{r={\rm const.}}$ of the spatial protection of the hypersurface.  An intersection of streamlines leads to $\left(\frac{\partial\theta}{\partial\theta_0}\right)=0$ at the point of intersection, which results in a divergence of the density at that point (see Eq. \eqref{density}). In this case the neglection of particle interaction is not a good approximation anymore. Therefor this approach can only be made if streamlines do not intersect, and Eq. \eqref{thth0c} holds. 
    
    To calculate the density field $n_p$ of a plasma with two types of test particles with different charges $e_1$ and $e_2$ and $e_1 e_2<0$, we simply compute 
    \begin{equation} n_p\left(r,\theta \right)=n_1\left(r,\theta\right)+n_2\left(r,\theta\right).\end{equation}
    Here $n_1$ and $n_2$ satisfy Eq. \eqref{density} for $e=e_1$ and $e=e_2$, respectively. By doing so, we assume that the particle densities of both types of test particles are sufficiently small, so that particle interactions are negligible.
   
   \subsection{The forming accretion disc} 
    As described in Sec. \ref{setup}, the particles from the spherical shell which do not fall onto the event horizon feed an initially featureless accretion disc located in the equatorial plane. We assume that in the disc particle interactions (viscosity, pressure, etc.) are not negligible anymore, and the particles that hit the accretion disc will be trapped in the disc.  Due to this process the accretion disc builds up until a stationary situation is reached. For the final form of the accretion disc we may then give up to two locations of very high densities (later called density peaks), and define the outer and the inner edge of the accretion disc as explained in the following.
   
    \paragraph*{The outer edge}
      We can define the outer edge of the forming accretion disc by bearing in mind relation \eqref{thth0c}. The furthest away a test particle with given initial conditions $\left( \dot r_0, \dot\phi_0, \dot\theta_0\right)$ can then reach the $\pi/2$ plane from the BH is given by 
       \begin{equation}
   	r_{\rm D}:= \lim_{\theta_0 \to \frac{\pi}{2}} r\left( \theta=\pi/2 \right)\,. %;E=E_{\frac{\pi}{2}},l=l_{\frac{\pi}{2}},K=K_{\frac{\pi}{2}} \right)
   	\label{rdeq}
       \end{equation}
       The point $r_{\rm D}$ then determines the outer edge of the final accretion disc. Note that $r(\theta)$, given by the Eqs. \eqref{rtheta} to \eqref{rmod}, depends on the roots of $\textbf{R}(r)$ and $\Theta(\theta)$ as well as on the constants of motion $E$, $l$, and $C$, which are all computed in the limit $\theta_0 \to \pi/2$ to determine $r_{\rm D}$.
       By using $\Psi(\theta_a)=0$ and $\Psi(\pi/2)=\frac{\cos\theta_a\,K(k_\theta)}{\sqrt{C+(E^2-1)a^2\cos\theta_a^4}}$, $K(k)$ being a complete elliptic integral of first kind, Eq. \eqref{xi} simplifies to
       \begin{align}
   	\xi_{\rm D} & = \frac{1}{2} \sqrt{\left(E^2-1\right)\left(r_a-r_c\right)\left(r_d-r_b\right)} \nonumber\\
   	& \quad \times \left[ \Phi\left(r_0\right)-
   	\frac{\pi}{2\sqrt{l^2-\left(E^2-1\right)a^2}}\right]~.
       \end{align}
       We again want to point out that here all quantities are computed in the limit $\theta_0 \to \pi/2$. 
       
     \paragraph*{The innermost stable circular orbit}
     \label{iscosec}
      Even though particle interactions are not negligible inside the accretion disc, we will use the assumptions of our dust accretion model to define the inner edge of the accretion disc. Since neither viscosity nor pressure occurs in our model, the inner edge of the accretion disc, which builds up due to the accretion process under discussion, will be located at the innermost stable circular orbit (ISCO) in Kerr--Newman spacetime. Particles that hit the equatorial plane at radii smaller than the ISCO are bound to fall into the black hole, and can therefore not contribute to the main accretion disc. However, they might form a so called minidisc, when spiraling into the black hole \citep{minidisc1, minidisc2}. We will come back to that, when discussing the accretion disc in Sec. \ref{results}.
      
      While the ISCO for Schwarzschild \citep{gravitation} is given by the simple expression of $r_{\rm ISCO} = 6M$, things are getting more complicated in Kerr spacetime. An exact expression for $r_{\rm ISCO}$ can still be derived \citep{lnrf}; however, two solutions arise for the ISCO in Kerr, one for direct and one for retrograde orbits. In Kerr spacetime the ISCO can reach from $M$ to $9M$ depending on the value of the rotation parameter. In Kerr--Newman spacetime one expects four different solutions for the ISCO in the case of charged particles. This can be traced back to the four possible combinations of direct or retrograde orbits and same charge ($eQ > 0$) or opposite charge ($eQ < 0$) of BH and test particles. As we neglect the magnetic monopole ($P = 0$), the accretion disc is located in the equatorial plane, and we are therefore interested in ISCOs for which $\theta=\pi/2$ holds.
  
       The ISCO is located where the effective potential of the $r$-motion $V_{\rm eff} (r) = \textbf{R}(r)$, see Eq. \eqref{dqr}, and its first and second derivative with respect to $r$ are equal to zero,
       \[ V_{\rm eff}(r_{\rm ISCO})=0~,~~{V'}_{\rm eff}(r_{\rm ISCO})=0~,~~{V''}_{\rm eff}(r_{\rm ISCO})=0~. \]
       Furthermore, as we are searching for ISCOs in the equatorial plane, the $\theta$-motion has to vanish at $\theta=\pi/2$, leading to
       \begin{equation} \left.\frac{d \theta}{d \lambda }\right|_{\theta=\pi/2}=0.\end{equation}
       Since we consider a very small charge of the BH as explained in Sec. \ref{setup}, it is for our purpose sufficient to solve the above equations for the ISCO for the case $Q = 0$ and $eQ \neq 0$. We find a complicated expression for $r_{\rm ISCO}$ (see appendix \ref{app2}), which can be solved numerically.  
       \begin{figure}
         \centering
	 \includegraphics{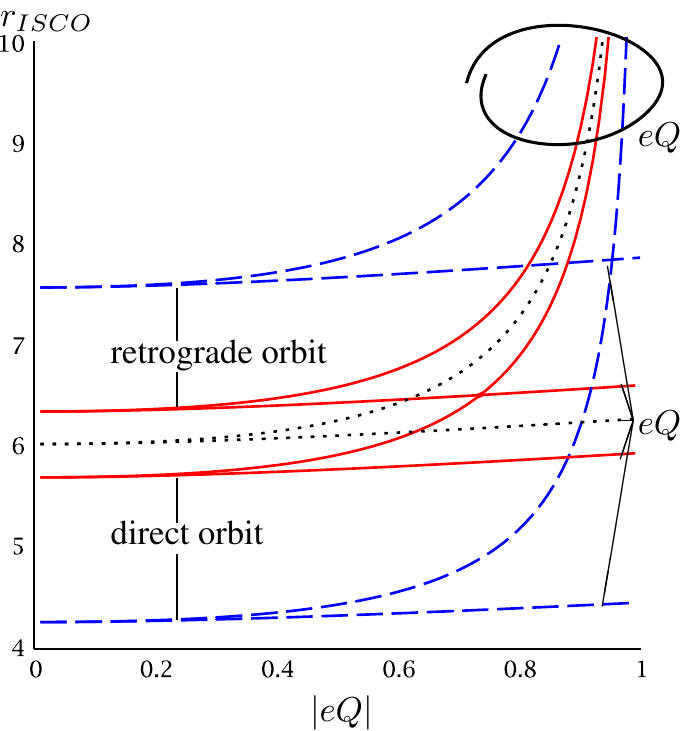}
	 \caption{Radius of the ISCO in the equatorial plane for charged particles in Kerr--Newman spacetime, with very small BH charge ($Q = 0$), for $a = 0$ (black, dotted), $a = 0.1$ (red, solid), and
   	       $a = 0.5$ (blue, dashed) as a function of $eQ$. Four different solution arise, traced back to the four combinations of direct or retrograde orbits and $eQ < 0$ or $eQ > 0$.
   	       The radius $r_{\rm ISCO}$ grows for bigger $\left|eQ\right|$ in all cases, but grows significantly faster for the case where BH and test particle have the same charge.}
        \label{iscoplot}
      \end{figure}
   
       The results are shown in Fig. \ref{iscoplot}. Four solutions for the ISCO can be found for each $eQ$ and $a \neq 0$. The black dotted curve represents the ISCO in Reissner--Nordstr\"om spacetime with vanishingly small $Q$. Starting from $r_{\rm ISCO} = 6M$ both solutions, for $eQ < 0$ and $eQ > 0$, grow for bigger values of $\left|eQ\right|$, causing $r_{\rm ISCO}$ to be minimal for uncharged particles, where $eQ=0$. While $r_{\rm ISCO}$ seems to grow somewhat exponentially for $eQ > 0$ it grows very slowly for $eQ < 0$. The same behavior can be seen for the ISCO in Kerr--Newman spacetime (red solid and blue dashed curves), but now four solutions arise. Two starting at each Kerr--ISCO for $\left|eQ\right| = 0$ and then show the same behavior for $eQ > 0$ and $eQ < 0$ with growing values of $\left|eQ\right|$ as in the Reissner--Nordstr\"om case.

\section{Results}
    \label{results}
      In this section we present solutions for plasma and uncharged dust accretion within the model described in Sec. \ref{setup}. For this, the streamlines, the three velocity field in LNRFs, and the density field are calculated for nine different combinations of the initial conditions and parameters (see Fig. \ref{sl1}--\ref{sl3}). Furthermore the influence of these on the value of the outer edge $r_{\rm D}$ (see definition in \eqref{rdeq}) of a forming accretion disc is discussed (see Fig. \ref{rd1} -- \ref{rd2}). For all plots in Fig. \ref{sl1}--\ref{sl3} the BH charge is chosen to be negative. The specific charge parameter of a proton and an electron will be called $e_p$ and $e_e$, respectively, in the following.
       
       We specify the initial conditions and parameters in the form  $(v^{\phi_0'}_{e}, v^{r_0'}_{e},Q,eQ,a)\in \left[0,1\right[$, where $v^{\phi_0'}_{e}$ and $v^{r_0'}_{e}$ are the radial and angular particle velocities in the LNRF at $r=r_0$ and 
       $\theta_0=\frac{\pi}{2}$ given in Eq. \eqref{vf1}--\eqref{vf3}. They have a one to one correspondence to $\dot r_0$, $\dot \phi_0$, and $\dot \theta_0$, which we choose as constant, in particular $\dot \theta_0=0$. 
       In case of a plasma the parameter $eQ$ is given for electrons. The parameter $eQ$ for protons is then 
       already determined and given by $e_pQ=\frac{\mu_e}{\mu_p}e_eQ$. The plotted solutions represent a family of solutions, since both the BH mass $M$ and the initial density $n_0$ at $r_0$ are not fixed.
    
      The density field, shown in Fig. \ref{sl1}--\ref{sl3}, is derived by numerically calculating the differential $\frac{\partial \theta}{\partial \theta_0}$ from Eq. \eqref{density}. Even though it is generally possible to derive an  analytical function for this derivative, $\theta(\theta_0)$ is a very complicated expression of $\theta_0$. It depends among others on the nodes of $\textbf{R}(r)$, which in turn depend on $\theta_0$ as well. We 
      refrain from calculating the derivative analytically and use a numerical method instead.
      
      \subsection{The velocity field, streamlines, and density field}
\begin{figure*}
	      \begin{minipage}{0.42\textwidth}
		      \includegraphics{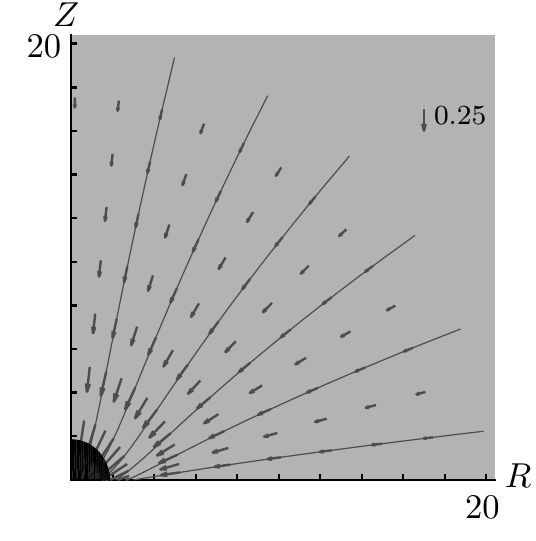}%[height=0.235\textheight]
		
		\vspace{0.6cm}
		(a) \hspace{0.2 cm} $e=0$, $a=0.1$, $Q=0.5$, $v^{\phi_0'}_{e}=0.11$, $v^{r_0'}_{e}=-0.2$
	      \end{minipage}
		\begin{minipage}{0.55\textwidth}
		      \includegraphics{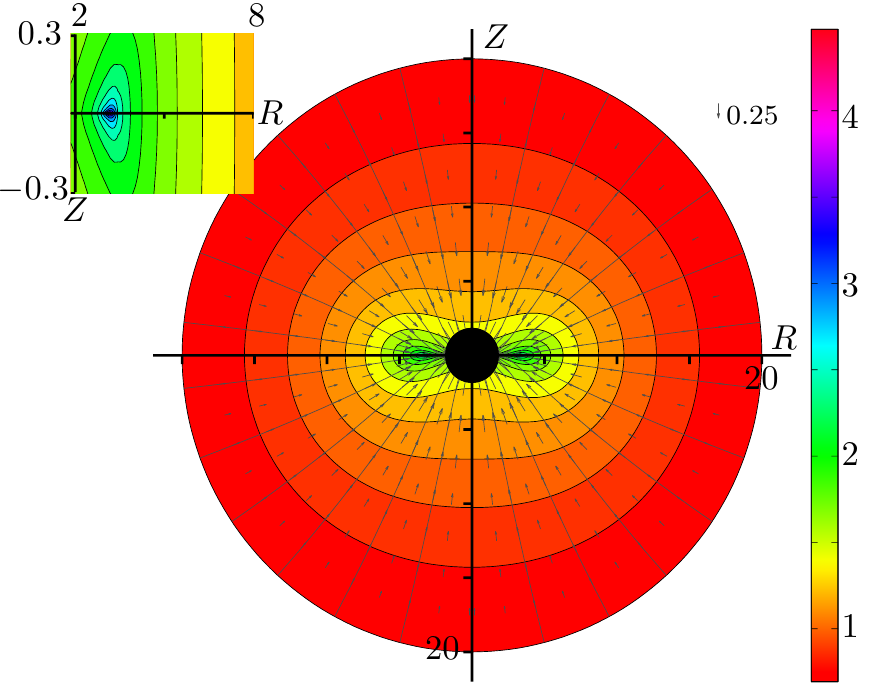}%[height=0.281\textheight]
	      \end{minipage}
	      
	      \vspace{0.5 cm}
	      \begin{minipage}{0.42\textwidth}
		    \includegraphics{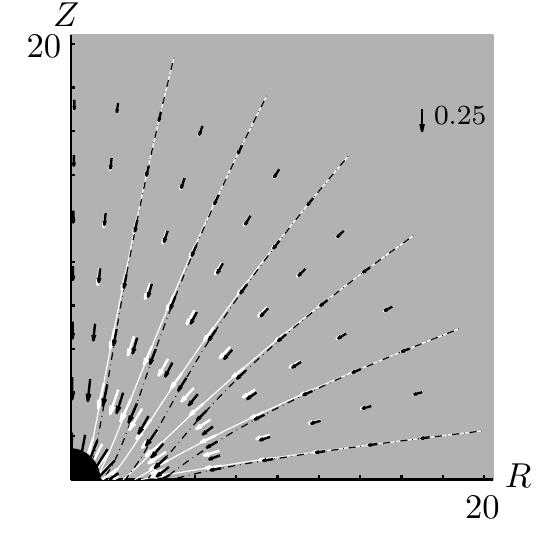}%[height=0.235\textheight]
		
		\vspace{0.6cm}
		(b) \hspace{0.2 cm} $a=0.9$, $e_eQ=0.5$ , $v^{\phi_0'}_{e}=0.11$, $v^{r_0'}_{e}=-0.2$(electrons)
	      \end{minipage}
	      \begin{minipage}{0.55\textwidth}
		      \includegraphics{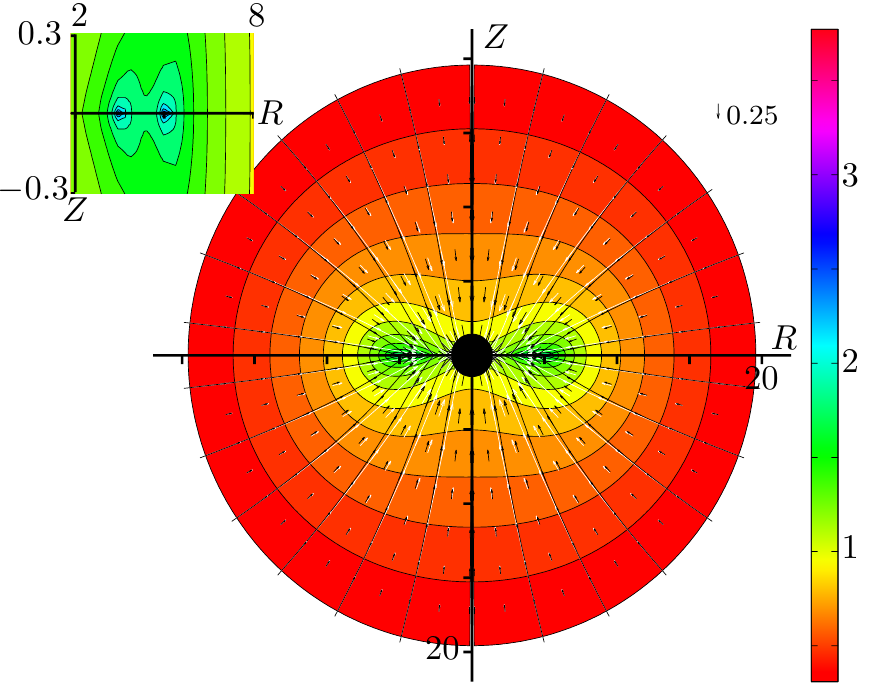}%[height=0.281\textheight]
	      \end{minipage}
	      
	      \vspace{0.5 cm}
	      \begin{minipage}{0.42\textwidth}
		      \includegraphics{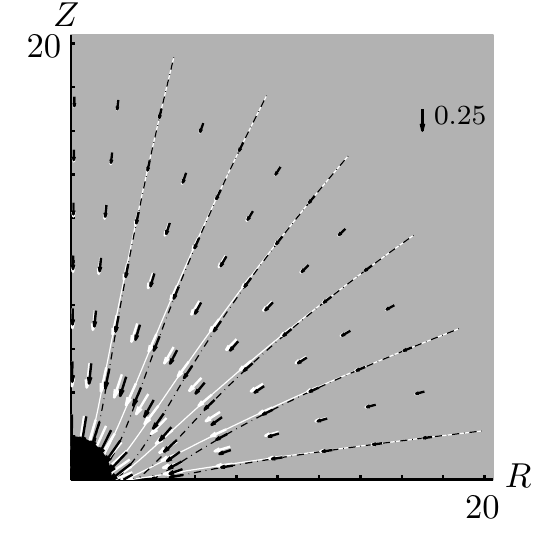}%[height=0.235\textheight]
		  
		\vspace{0.6cm}
		(c) \hspace{0.2 cm} $a=0.1$, $e_eQ=0.5$, $v^{\phi_0'}_{e}=0.11$, $v^{r_0'}_{e}=-0.2$ (electrons)
	      \end{minipage}
		\begin{minipage}{0.55\textwidth}
		      \includegraphics{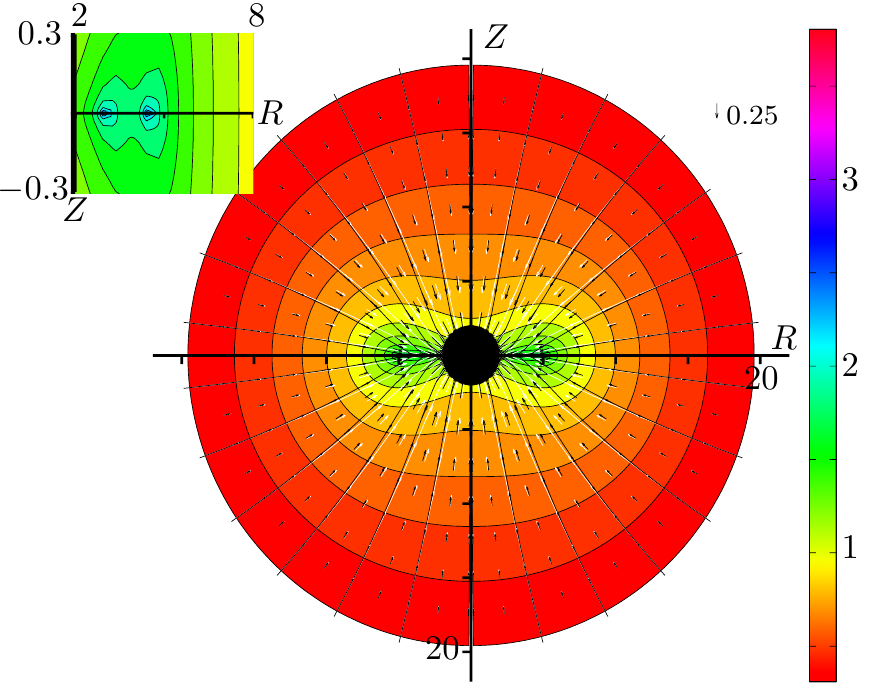}%[height=0.281\textheight]
	      \end{minipage}
	      \caption{ Streamlines, three velocity field in LNRFs (left and right) and density field (left) are plotted for plasma or neutral particles, a negatively charged BH, $r_0=20$ and different initial conditions.
	      The corresponding parameter $eQ$ for protons can be calculated by $e_pQ=\frac{\mu_e}{\mu_p}e_eQ$. Black, white and gray streamlines and velocity fields describe electron, proton and neutral particle motion respectively. The density color bar is given in a logarithmic scale. The initial condition $v^{x}_{e}$ is the $x$-component of the three velocity at $r=r_0$ and $\theta=\pi/2$, given by Eqs. \eqref{vf1}--\eqref{vf3}. Two density peaks arise, which can be traced back to the two differently charged particle types of the plasma. Changes in the initial conditions $v^{\phi_0'}_{e}$ and $v^{r_0'}_{e}$ and $e_eQ$ have a strong effect on all features of the accretion flow. This effect of the initial velocities and $e_eQ$ can be studied by comparing the plots from Figs. \ref{sl1}-- \ref{sl3} with each other.}%of the streamlines, the velocity field and the density field.}
	      \label{sl1}
      \end{figure*}

      Figure \ref{sl1} (a) shows the special accretion case of uncharged ($e=0$) dust on a strongly charged BH ($Q\approx1$). Since for $e=0$ the BH charge $Q$ only appears in $\Delta(r)$, its influence on the accretion flow is non-negligible only close to the horizon, where  $\Delta(r)$ approaches zero (see Eq. \eqref{dqr}, \eqref{dqth} and \eqref{delta}). This statement is supported by comparing the results for the outer edge in plot \ref{rd2} (a) with the ones in \ref{rd1} (a) and (b). In plot \ref{rd2} (a) it can be seen that the increase of $r_{\rm D}$ from $Q=0$ to $Q=\sqrt{1-a^2}$ is of the order of $10$ per cent. This is rather small compared to the increase of $r_{\rm D}$ caused by a change of the initial conditions $v^{r_0'}_{e}$ and $v^{\phi_0'}_{e}$, shown in plot \ref{rd1} (a) and (b) at $eQ=0$ respectively. Here the value of  $r_{\rm D}$ might even more then double. Over all the accretion flow for $e=0$, shown in Fig. \ref{sl1} (a), approaches the one for the Kerr spacetime, discussed in \cite{acck}, and is therefore mainly given as an example for uncharged particle accretion.
  
\begin{figure*}
		\begin{minipage}{0.42\textwidth}
		      \includegraphics{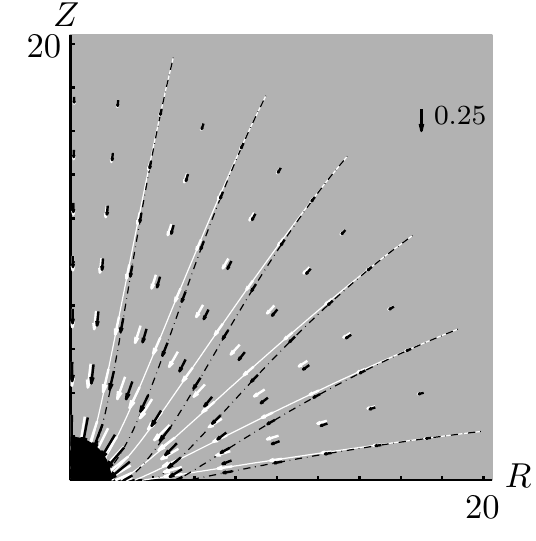}%[height=0.235\textheight]
		  
		\vspace{0.6cm}
		(a) \hspace{0.2cm} $a=0.1$, $e_eQ=0.5$, $v^{\phi_0'}_{e}=0.11$, $v^{r_0'}_{e}=-0.1$ (electrons)
	      \end{minipage}
	      \begin{minipage}{0.55\textwidth}
		      \includegraphics{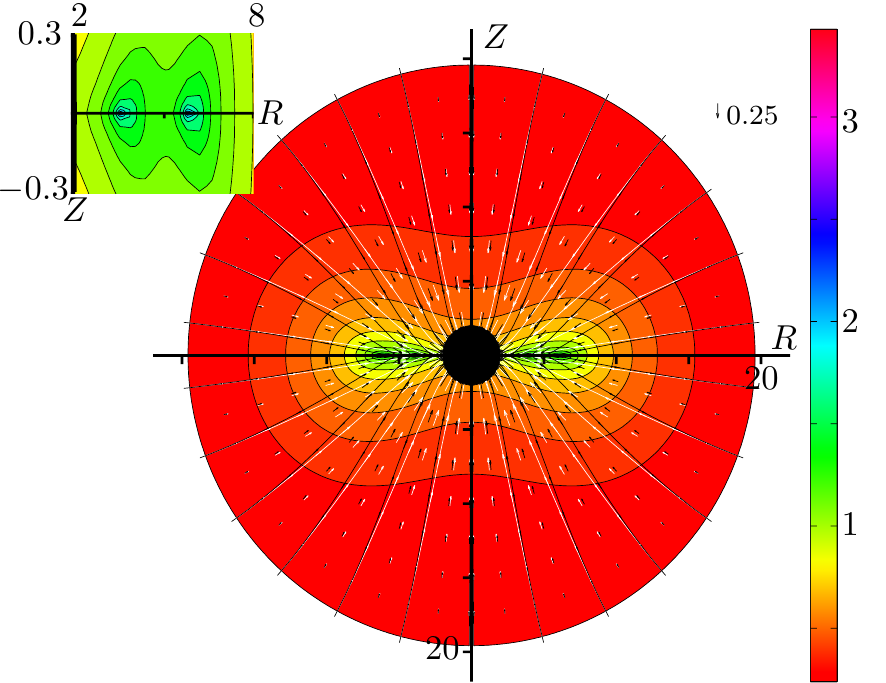}%[height=0.281\textheight]
	      \end{minipage}
	      
	      \vspace{0.5 cm}
	      \begin{minipage}{0.42\textwidth}
		      \includegraphics{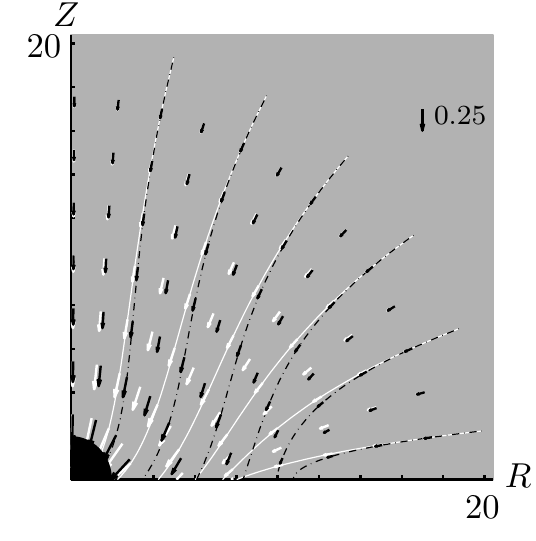}%[height=0.235\textheight]
		  
		  \vspace{0.6cm}
		  (b) \hspace{0.2 cm}  $a=0.1$, $e_eQ=0.5$, $v^{\phi_0'}_{e}=0.2$, $v^{r_0'}_{e}=-0.2$ (electrons)
	      \end{minipage}
	      \begin{minipage}{0.55\textwidth}
		      \includegraphics{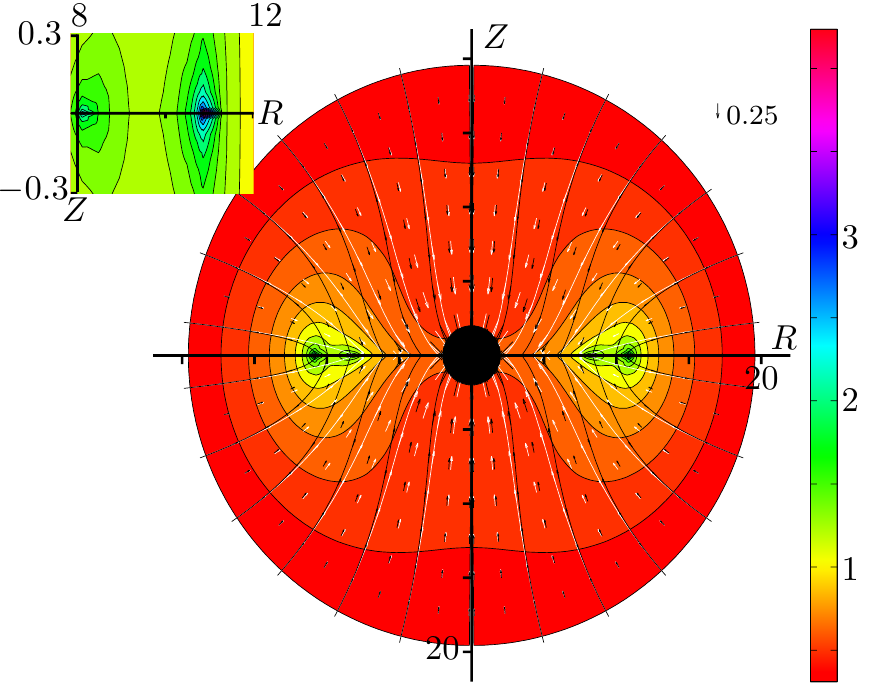}%[height=0.281\textheight]
	      \end{minipage}
	      
	      \vspace{0.5 cm}
		\begin{minipage}{0.42\textwidth}
		      \includegraphics{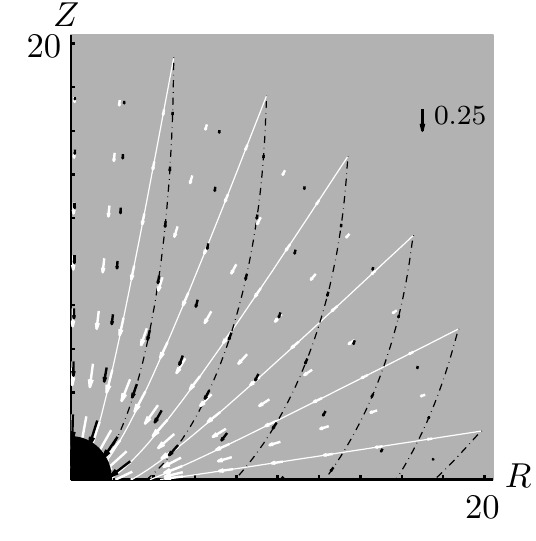}%[height=0.235\textheight]

		  \vspace{0.6cm}
		(c) \hspace{0.2 cm} $a=0.1$, $e_eQ=0.8$, $v^{\phi_0'}_{e}=0.11$, $v^{r_0'}_{e}=-0.001$
		
		      \hspace{0.6cm} (electrons)
	      \end{minipage}
		\begin{minipage}{0.55\textwidth}
		      \includegraphics{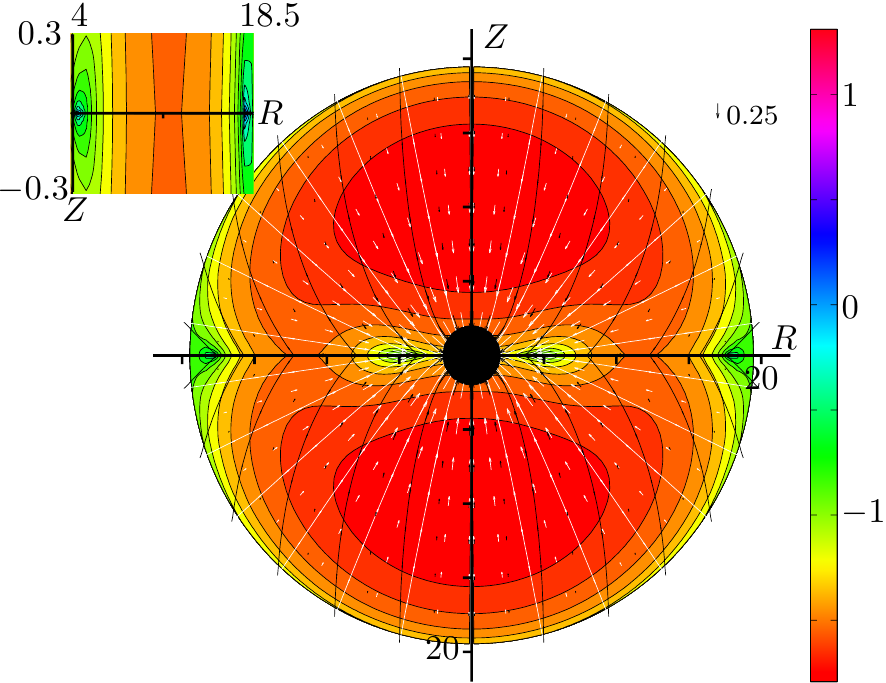}%[height=0.281\textheight]
		\end{minipage}
	      \caption{For a detailed description see caption of Fig. \ref{sl1}. A comparison of the plot (a) and (b) with the plot in Fig. \ref{sl1} (c) shows the influence of the initial velocities $v^{r_0'}_{e}$ and $v^{\phi_0'}_{e}$ respectively. In plot (c) the initial value for the $r$-motion is chosen to be very small. This results in a weak particle infall, leading to very small density values and a big effect of the attractive and repulsive electromagnetic force on the accretion flow.}
	      \label{sl2}
      \end{figure*}

\begin{figure*}
	    \begin{minipage}{0.42\textwidth}
		    \includegraphics{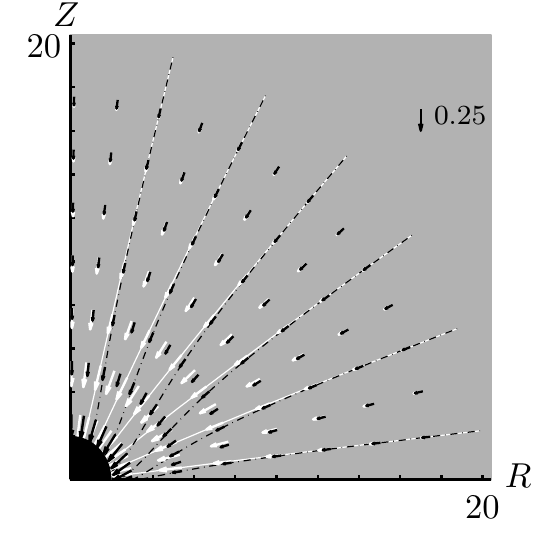}%[height=0.235\textheight]
		
	      \vspace{0.6cm}
	      (a) \hspace{0.2 cm} $a=0.1$, $e_eQ=1$, $v^{\phi_0'}_{e}=0.06$, $v^{r_0'}_{e}=-0.2$ (electrons)
	    \end{minipage}
	    \begin{minipage}{0.55\textwidth}
		    \includegraphics{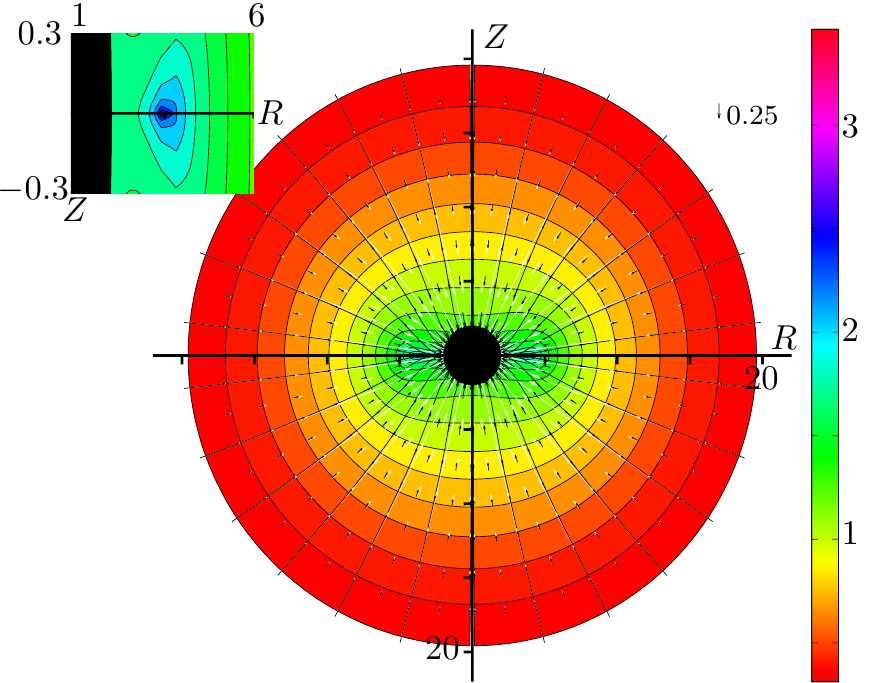}%[height=0.281\textheight]
	    \end{minipage}
	    
	    \vspace{0.5 cm}
	    \begin{minipage}{0.42\textwidth}
		    \includegraphics{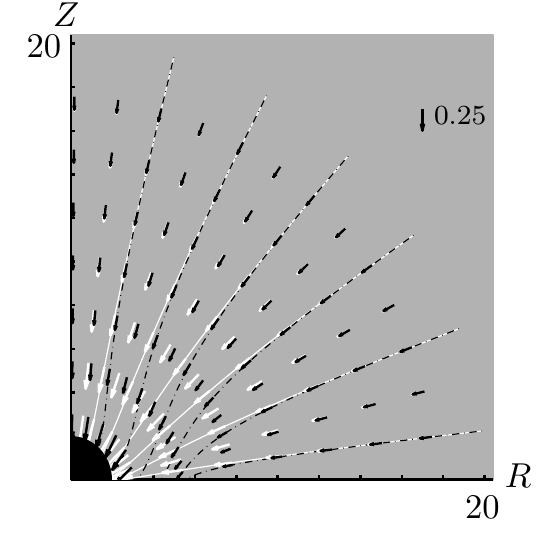}%[height=0.235\textheight]
		
		\vspace{0.6cm}
	      (b) \hspace{0.2 cm} $a=0.1$, $e_eQ=1$, $v^{\phi_0'}_{e}=0.11$, $v^{r_0'}_{e}=-0.3$ (electrons)
	    \end{minipage}
	    \begin{minipage}{0.55\textwidth}
		    \includegraphics{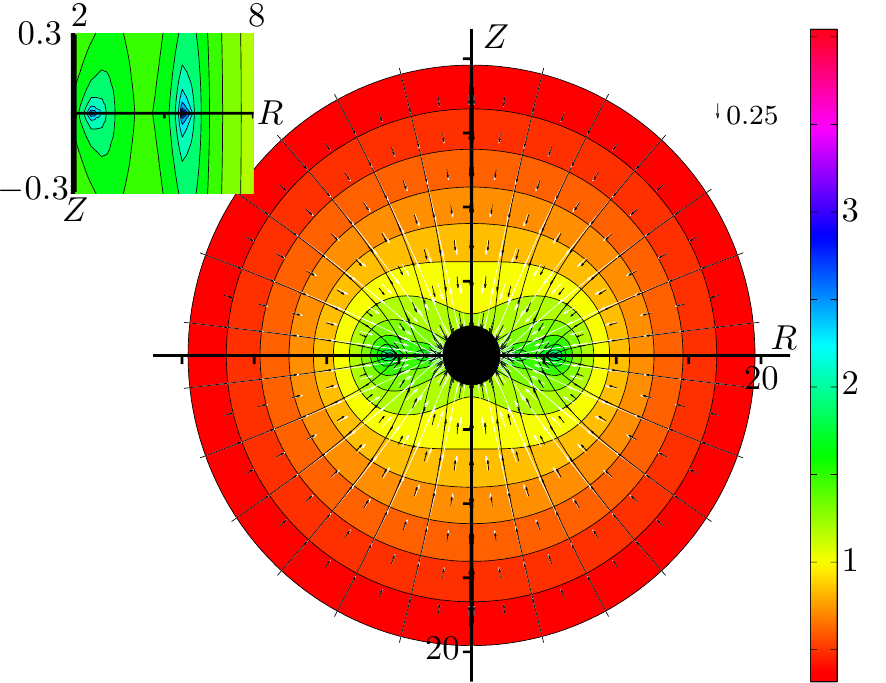}%[height=0.281\textheight]
	    \end{minipage}
	    
	    \vspace{0.5 cm}
	    \begin{minipage}{0.42\textwidth}
		    \includegraphics{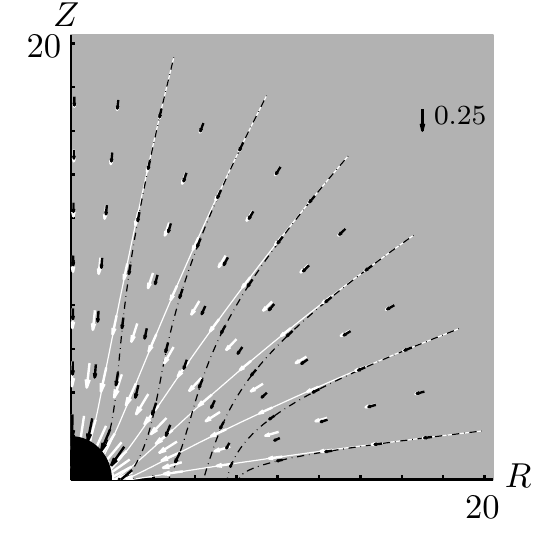}%[height=0.235\textheight]
		
	      \vspace{0.6cm}
	      (c) \hspace{0.2 cm} $a=0.1$, $e_eQ=1$, $v^{\phi_0'}_{e}=0.11$, $v^{r_0'}_{e}=-0.2$(electrons)
	    \end{minipage}
	    \begin{minipage}{0.55\textwidth}
		    \includegraphics{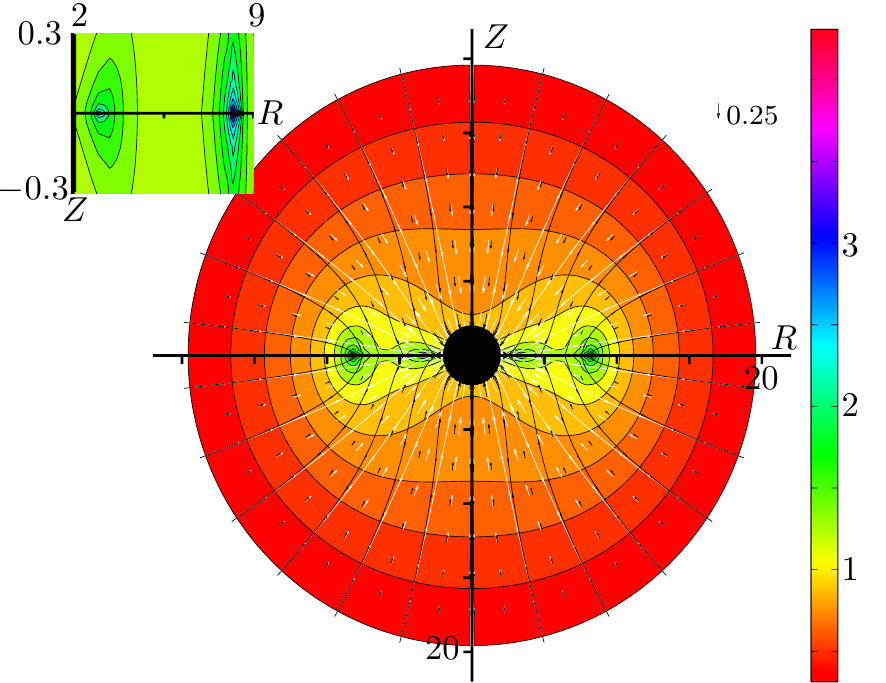}%[height=0.281\textheight]
	    \end{minipage}
	    \caption{ For a detailed description see caption of Fig. \ref{sl1}. A comparison of plot (a) with plot (c) and a comparison of plot (b) with plot (c) show the influence of the initial velocities $v^{\phi_0'}_{e}$ and $v^{r_0'}_{e}$ respectively. The influence of $e_eQ$ on the accretion flow is shown by a comparison with plot (c) and the plot in Fig. \ref{sl1} (c). In plot (a) only one density peak arises, produced by the accreted electrons. All streamlines of the proton accretion flow reach the BH horizon before hitting the $\theta=\pi/2$ plane and therefore will not create a density peak.}
	    \label{sl3}
      \end{figure*}

      Taking a look at Figs. \ref{sl1} -- \ref{sl3} we see that for a plasma two density peaks will arise at the $\pi/2$ plane. They are each caused by one of the two different particle types (distinguished by their different specific charges $e_p$ and $e_e$). Furthermore the influence of the metric's parameters $a$, $eQ$ and the initial velocities $v^{\phi_0'}_{e}$ and $v^{r_0'}_{e}$ on the accretion flow and the position of the density peaks is pictured in this figures. This will be discussed in more detail in the following.
      
      The plots in Fig. \ref{sl1} (b) and (c) show the accretion flow for the same initial conditions but different angular momenta $a$. These plots are given as an example to show that a variation of $a$ does only weakly effect the accretion flows onto the BH. The overall structure of the accretion flow stays the same, while only a small shift in the position of the density peaks can be detected. A more detailed discussion of the influence of $a$ onto the accretion flow was done in \cite{acck}, which is why we won't go into further details here.
      
      The influence of the initial conditions and parameters in the model can be analyzed by comparing plots where only one of the parameters $v^{\phi_0'}_{e}$, $v^{r_0'}_{e}$ or $eQ$ is changed. We first analyze the influence of $v^{\phi_0'}_{e}$ on the accretion flow by comparing Fig. \ref{sl1} (c) with \ref{sl2} (b) and Fig. \ref{sl3} (a) with \ref{sl3} (c). This shows that the bigger the value of $v^{\phi_0'}_{e}$ the stronger the course of the streamlines deviate from a radial infall. The same statement holds if we analyze the influence of $v^{r_0'}_{e}$ on the accretion by comparing Fig. \ref{sl1} (c) with \ref{sl2} (a) and Fig. \ref{sl3} (b) with \ref{sl3} (c); also, this is true for the influence of $eQ$ which can be seen by comparing Fig. \ref{sl1} (c) with \ref{sl3} (c). Summarized, the larger we choose $v^{\phi_0'}_{e}$, $v^{r_0'}_{e}$ or $eQ$, the stronger curved the streamlines are. This is also why for a negatively charged central BH the course of electrons is stronger influenced than the course of protons, since $e_pQ<<e_eQ$. 
      
      Figure \ref{sl2} (c) shows the biggest difference between the particle flow of the two different particle types. Here 
      the initial value for the $r$-motion with $v^{r_0'}_{e}=-0.001$ is chosen very small. As a result there is a very weak particle infall, leading to very small density values (see Eq. \eqref{density}). On the other hand, since 
      the initial $r$ velocity of the infalling particles is very slow, the attractive and repulsive electromagnetic force on the particles show more effect on their course. While the streamlines of attracted particles (white lines) show a close to radial infall, the streamlines of the repulsed particles (black lines) show the typical course of a small value of $d\,r/d\,\theta=\sqrt{\textbf{R}\left(r\right)/\Theta\left(\theta\right)}$. This arises from the fact that $d\,r/d\,\lambda=\sqrt{\textbf{R}\left(r\right)}$ stays small in case of a repulsive electromagnetic force.
    
      \subsection{The accretion disc}
\begin{figure}
	   \includegraphics{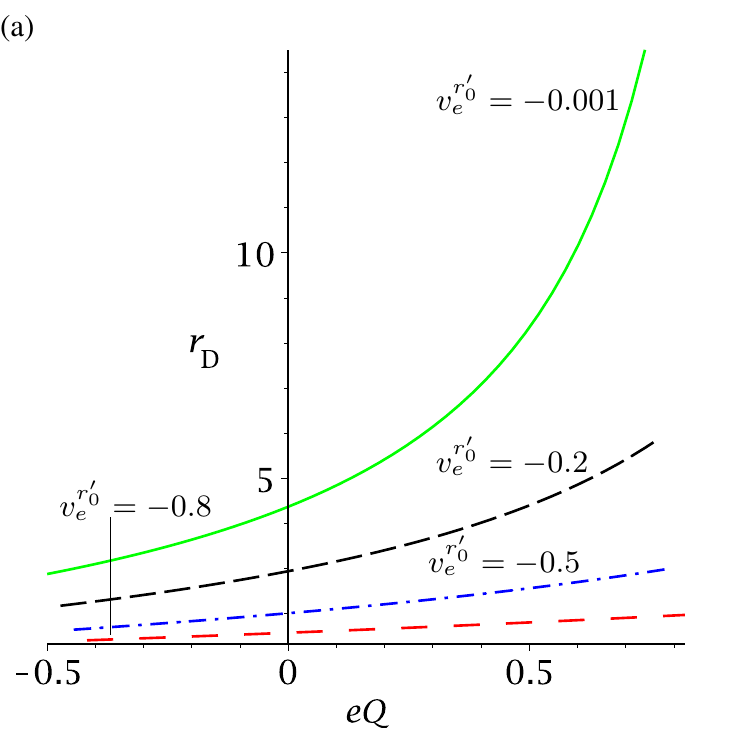}
	   \includegraphics{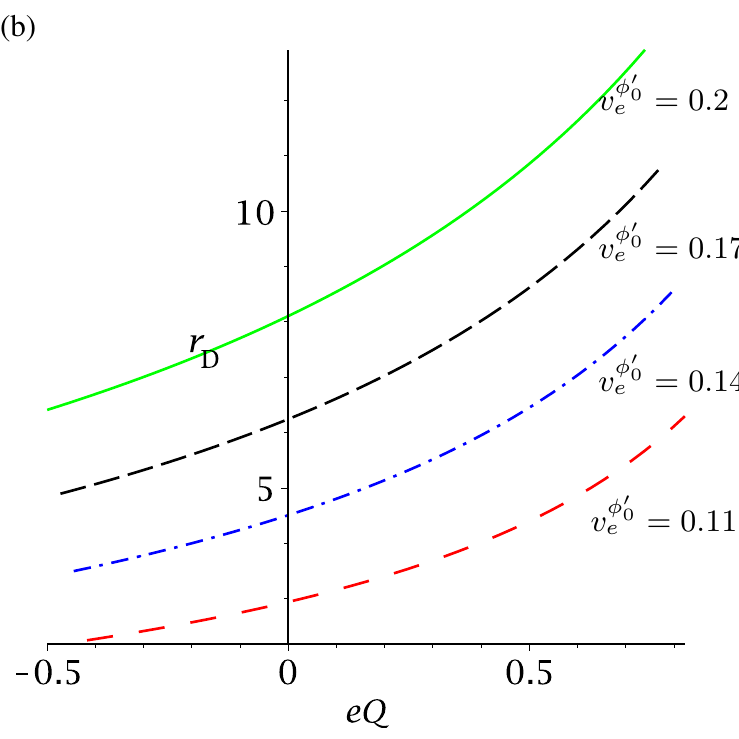}
	   \caption{Outer edge $r_{\rm D}$ of the forming accretion disc as a function of $eQ$ for $r_0=20$ and $a=0.1$. Here $v^{x}_{e}$ gives the $x$ component of the three velocity at $r=r_0$, $\theta=\pi/2$. (a) $v^{\phi_0'}_{e}=0.11$, different $v^{r_0'}_{e}$. (b) $v^{r_0'}_{e}=-0.2$, different $v^{\phi_0'}_{e}$. The dependence of $r_{\rm D}$ on $eQ$ increases with growing values of $v^{\phi_0'}_{e}$ and $v^{r_0'}_{e}$.}
         \label{rd1}
      \end{figure}

\begin{figure}[h!]
	\center{\includegraphics{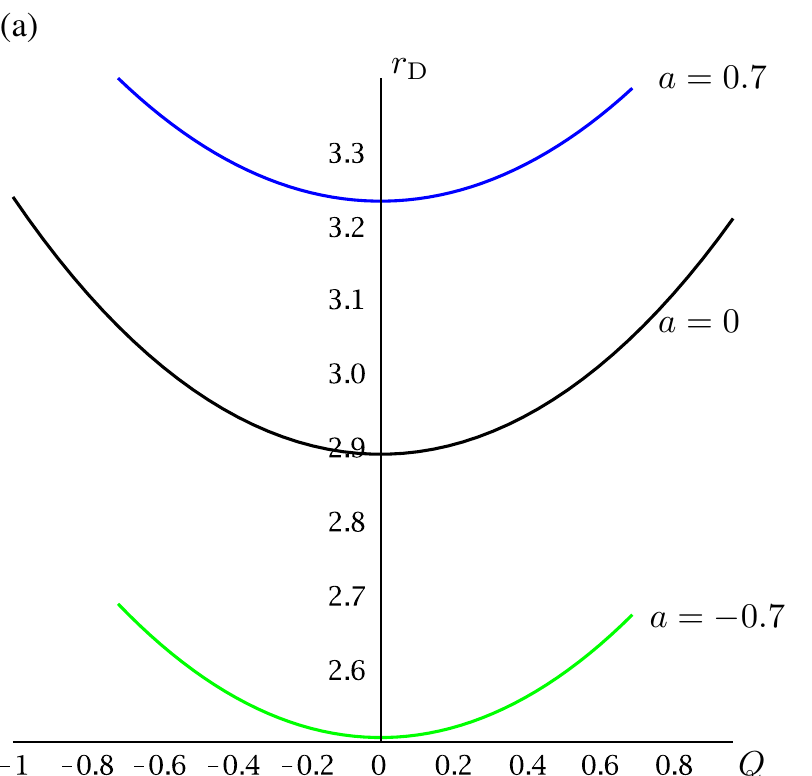}}
        \center{\includegraphics{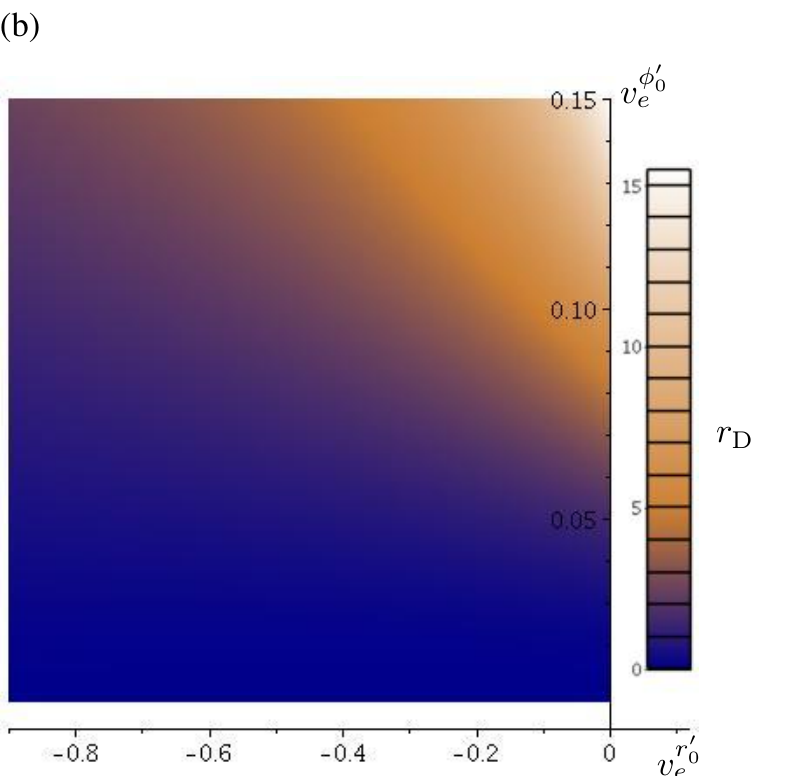}}
        \caption{Outer edge $r_{\rm D}$ of the forming accretion disc as (a) a function of $Q$ for $r_0=20$, $e=0$, $v^{\phi_0'}_{e}=0.11$, $v^{r_0'}_{e}=-0.2$ and different values for $a$ and (b) a function of  $v^{\phi_0'}_{e}$ and $v^{r_0'}_{e}$ for $r_0=20$, $e_eQ=0.5$, $a=0.1$. The dust flow is counter-rotating for $a<0$ and corotating for $a>0$. It can be seen in plot (a), that $r_{\rm D}$ changes only slightly with variation of $a$, even less with variation of $Q$ compared to the changes induced by a variation of the initial velocities $v^{\phi_0'}_{e}$ and $v^{r_0'}_{e}$, plotted in (b). This changes become bigger for  bigger values of $v^{\phi_0'}_{e}$ and $v^{r_0'}_{e}$. The shift of $r_{\rm D}$ for growing $a$ to bigger values depicts the frame-dragging effect.}
       \label{rd2}
      \end{figure}

      Figure \ref{rd1} and Fig \ref{rd2} (b) picture the influence of $eQ$ and the initial velocities of the test particles on the position of the outer edge $r_{\rm D}$. From this we can conclude that $\frac{d r_{\rm D}}{dp}$, where $p$ is one of the parameters $v^{\phi_0'}_{e}$, $v^{r_0'}_{e}$ or $eQ$, is largest for big values of the parameters. Therefore, the influence of a small change in one of the parameters $v^{\phi_0'}_{e}$, $v^{r_0'}_{e}$ or $eQ$ is rather small if the parameter is small, but becomes significant for bigger values of the parameters, see table \ref{table1}.
      \begin{table}[t]
	\begin{center}
	  \begin{tabular*}{7.8 cm}{@{\extracolsep{\fill} } l c c}
	    \hline \hline
						& Example 1 & Example 2 \\
	    \hline
	    $v^{\phi_0'}_{e}$			 & $0.11$    & $0.11$    \\
	    $v^{r_0'}_{e}$			 &$-0.27$    & $-0.001$	 \\
	    $eQ_1$				 &$-0.5$     & $0.5$ 	 \\
	    $eQ_2$			 	 &$0$	     & $0.7$ 	 \\
	    $\Delta r_{\rm D} (eQ_1\rightarrow eQ_2)$&$\approx 0.5$& $\approx 6$\\
	    \hline \hline
	  \end{tabular*}    
	\end{center}
	\caption{Comparison of the change of the outer edge $\Delta r_{\rm D}$ for an increase of $eQ$ from $eQ_1$ to $eQ_2$ between two sets (example 1 and example 2) of initial conditions and $eQ$. The influence of changing the value of $eQ$ results in a significant bigger change of $r_{\rm D}$ in example 2, where bigger values for the initial conditions and $eQ$ are chosen.}
	\label{table1}
      \end{table}
    
      The outer edge of the disc does not depend much on the specific electric charge $Q$ of the BH  (see Fig. \ref{rd2} (a)), as already discussed before. The influence of the angular momentum $a$ of the BH (see Fig. \ref{rd2} (a)) on $r_{\rm D}$ is also small. (As an example, consider for $e=0$, $Q=0$, $v^{r_0'}_{e}=-0.2$, $v^{\phi'}_{e}=0.112$ the increase $\Delta r_{\rm D}(a_1\rightarrow a_2) \approx 0.7$, for a change of $a$ from $a_1=-0.7$ to $a_2=0.7$). However, the shift of the outer edge of the disc for different values of $a$ due to the frame-dragging effect, which was already stressed in \cite{acck}, is reproduced here. A counter-rotating flow ($a<0$) leads to a smaller value of the outer egde $r_{\rm D}$ as compared to the the corotating case ($a>0$). 
             
      If we identify the two density peaks as the positions of the outer edges of particle type $1$ and $2$, which we conclude from the examples presented here, we infer from the above discussion that the distance between the density peaks weakly depends on the parameters $Q$ and $a$, but strongly depends on the product $eQ$ and the initial velocities $v^{\phi_0'}_{e}$ and $v^{r_0'}_{e}$ for sufficiently big values of these parameters. The distance grows for increasing values of $eQ$, $v^{\phi_0'}_{e}$ and $v^{r_0'}_{e}$.
    
      In case of a plasma we can calculate two different values for both $r_{\rm D}$ and the ISCO for given initial conditions, one due to the electrons and one due to the protons forming the plasma. The inner and outer edge of the formed accretion disc should then be defined by one of the two values for the ISCO and one of the two values for $r_{\rm D}$ respectively. Minidiscs might build up for radii smaller than the inner edge of the main accretion disc, where the matter is bound to spiral into the BH. Particles hitting the equatorial plane for radii larger than $r_{\rm{ISCO}}$ may in principle loose so much energy that they, as well, are bound to spiral into the BH, forming a minidisc. However, at this point the interaction with the main accretion disc should not be neglected, and the model description breaks down. We will therefore not further discuss this possibility here.
      
      Within this setting four cases can occur for a plasma, which we discuss below. Here we indicate the specific charge of particles with the opposite charge of the BH with $e_1$ ($e_1Q<0$), and the specific charge of particles with the same charge as that of the BH with $e_2$ ($e_2Q>0$).
			
      \begin{figure}
	\centering
	\includegraphics{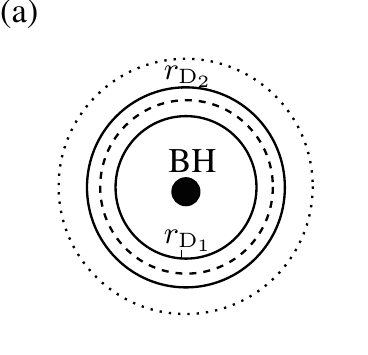}
	\includegraphics{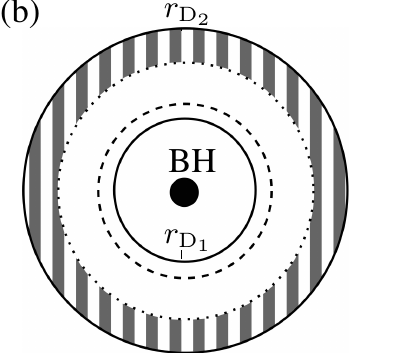}

	\includegraphics{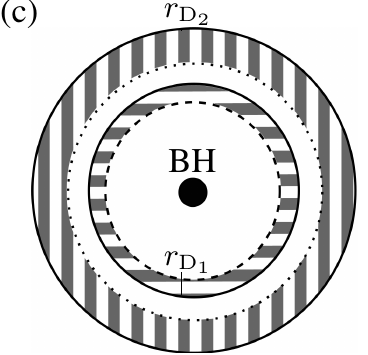}
	\includegraphics{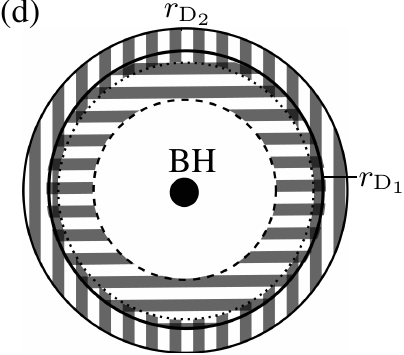}
	
	\includegraphics{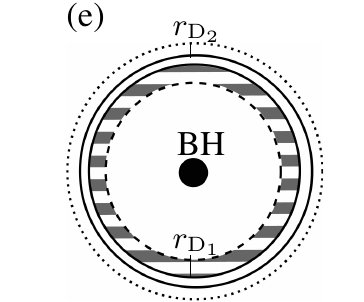}
	\includegraphics{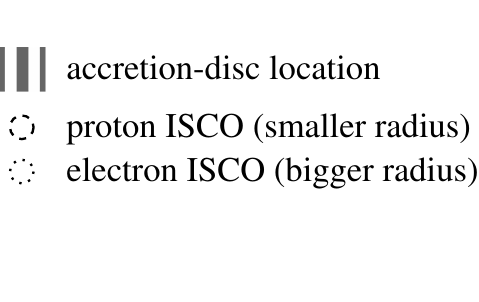}
	
	\caption{Schematic plots of possible accretion disc scenarios. Here ${r_{\rm D_1}}=r_{\rm D}\big|_{e_1Q}$ corresponds to $e_1Q<0$ and ${r_{\rm D_2}}=r_{\rm D}\big|_{e_2Q}$. If the outer edge $r_{\rm D}$ is smaller than the corresponding ISCO, no accretion disc is formed.  For a general description see Case 1 to Case 4 in Sec. \ref{results}, where (a) corresponds to Case 1, (b) corresponds to Case 2, (c) and (d) correspond to Case 3, and (e) corresponds to Case 4.}
	\label{shematicplot}
      \end{figure}
      \paragraph*{Case 1} $r_{\rm ISCO}\big|_{e_1Q}>r_{\rm D}\big|_{e_1Q}$ and $r_{\rm ISCO}\big|_{e_2Q}>r_{\rm D}\big|_{e_2Q}$:
    
	      All matter reaches the $\pi/2$ plane for radii smaller than the ISCO (see schematic plot in Fig. \ref{shematicplot} (a)). All accreted matter is bound to spiral into the BH and might form a minidisc during this process. It will not contribute to or form a main accretion disc. This case occurs for small enough $v^{r_0'}_{e}$ and $v^{\phi_0'}_{e}$, e.g.~for a negatively charged BH, where  $e_2Q=0.5$ (electrons), $e_1Q=-0.00027$ (protons) and all values for $v^{r_0'}_{e}$ and $v^{\phi_0'}_{e}$ where  $r_{\rm D} \lesssim 5.5$ (compare Fig. \ref{rd2} (b))
      \paragraph*{Case 2} $r_{\rm ISCO}\big|_{e_1Q}>r_{\rm D}\big|_{e_1Q}$ and $r_{\rm ISCO}\big|_{e_2Q}<r_{\rm D}\big|_{e_2Q}$:
      
	      All particles with a charge opposite to the BH spiral into it. But the majority of streamlines of particles with the same charge as the BH will reach the $\pi/2$ plane for radii bigger than the corresponding ISCO, since the density peak is located at $r_{\rm D}$, and therefore contribute to or form an accretion disc. In this situation the accretion disc should slowly develop the same charge as the BH  (see schematic plot in Fig. \ref{shematicplot} (b)), until the electromagnetic field created by the disc's charge is not negligible anymore and the model's description breaks down. This case occurs for example for a negatively charged BH, where $e_2Q=0.5$, $e_1Q=-0.00027$, $v^{r_0'}_{e}=-0.2$ and $v^{\phi_0'}_{e}=0.13$.
      \paragraph*{Case 3} $r_{\rm ISCO}\big|_{e_1Q}<r_{\rm D}\big|_{e_1Q}$ and $r_{\rm ISCO}\big|_{e_2Q}<r_{\rm D}\big|_{e_2Q}$:
      
	      The majority of the streamlines of all particles reach the $\pi/2$ plane for radii bigger than the corresponding ISCO (see schematic plot in Fig. \ref{shematicplot} (c), (d)). Since $r_{\rm D}$ and the ISCO are smaller for particles which have a charge opposite to that of the BH than for those whose charge has the same sign as the BH, within the model's description we expect an inner and outer area of the accretion disc. Here the inner area is dominated by oppositely charged particles ($e_1Q<0$), and the outer area is dominated by particles of the same charge ($e_2Q>0$). However interactions between the particles should not be neglected at the accretion disc and interactions might prevent the development of these areas within the accretion disc. Like in Case 2, the model description might beak down for this case, if the electromagnetic field created by the oppositely charged areas can not be neglected anymore. This case occurs for sufficiently big values for $v^{r_0'}_{e}$ and $v^{\phi_0'}_{e}$, e.g.~for a negatively BH, where $e_2Q=0.5$, $e_1Q=-0.00027$, $v^{r_0'}_{e}=-0.2$ and $v^{\phi_0'}_{e}\geq 0.17$.
      \paragraph*{Case 4} $r_{\rm ISCO}\big|_{e_1Q}<r_{\rm D}\big|_{e_1Q}$ and $r_{\rm ISCO}\big|_{e_2Q}>r_{\rm D}\big|_{e_2Q}$:
      
	      All particles with the same charge as the BH spiral into it, while oppositely charged particles can stay on the $\pi/2$ plane (see schematic plot in Fig. \ref{shematicplot} (e)). This would be a situation where the accretion disc slowly develops a charge, opposite to the BH's charge. This case might occur for very big values of $e_eQ\rightarrow 1$ and sufficiently big values of $v^{r_0'}_{e}$ and $v^{\phi_0'}_{e}$. However, since always ${r_{\rm D_2}}>{r_{\rm D_1}}$, while spiraling inwards the particles of the same charge as the BH will have to pass through the area where the model predicts an accumulation of oppositely charged particles. We have to expect interactions between the particles at this point and the models prescription breaks down. These interactions probably prevent the oppositely charged particles to actually fall into the BH. An accretion disc slowly developing a charge with the same sign as the BH therefore is an interesting but unlikely scenario.
	      
      
  \subsection{Limits of the model due to electromagnetic particle interactions}
      The negligence of particle interactions, especially the electromagnetic interactions, of the used model has its limits. For a plasma the model predicts the occurrence of two local density maxima, one for each particle type, with a very sharp density peak at its center. At this center the electromagnetic particle interactions will most likely not be negligible anymore. The occurring repulsive electromagnetic forces at these points will have the effect of softening the sharp peaks. However, since these sharp peaks lie on the equatorial plane, they will further, and probably much more strongly, be effected by the accretion disc, which is assumed to be located there as well.
      
      Neglectring particle interactions also restricts our choice in the initial particle density $n_0$ at $\bar r_0$. The electromagnetic field, created by the infalling plasma particles, should still be negligible compared to the field created by the BH. As a result, a limit for $n_0$ depends on the choice of $\bar r_0$, the total BH mass $M$, and on the position of the density peaks, which create the electromagnetic field of up to two charged rings around the BH. Changing $\bar r_0$ to bigger values raises the strength of the electromagnetic field of the infalling particles at the outer area, while at the same time the BH's electromagnetic field falls off. Therefore $n_0$ has to be chosen to be smaller for larger values of $\bar r_0$. The same holds for the value of the BH mass $M$ for constant $\bar r_0=\frac{r_0}{M}$, since the total charge of the accreted particles scales with $M^3$, while the total charge $Q=M\,\bar Q$ of the BH scales with $M$. Furthermore $n_0$ has to be chosen to be smaller the further the density peaks are located away from the BH. This results from the same consideration done for the effect of $\bar r_0$.
      
      For a stellar BH with a net charge of $10^{-18}$ -- $10^{-21}$ the plasma density is restricted to values smaller than $10$--$0.01\,\rm{cm}^{-3}$. The density has to be even more dilute for more massive BHs or bigger chosen values for $\bar r_0$. 
      
      In this subsection we came back to the notation of Sec. \ref{ecom}, and wrote the net charge and radius with bars, where it is given in it's dimensionless form. 
      
\section{Summary and Conclusion \label{conclusion}}
    We discussed an analytical model for the relativistic accretion of (charged) dust onto a rotating and charged black hole as described by the Kerr--Newman spacetime. Our model is a direct generalization of the papers by 
    %E. Tejeda, P. Taylor and J. C. Miller 
    \citet*{accss} and \citet*{acck} on dust accretion onto a Schwarzschild and Kerr BH. Because strongly charged BHs are astrophysically quite unlikely, we assumed here very weakly charged BHs with a normalized charge  parameter $Q$ of the order of $10^{-18}$ -- $10^{-21}$. However, for either electrons or protons with a normalised charge $e$ we may then find $eQ \approx 1$, which results in a quite significant influence on the accretion process. In our streamline and density plots, however, we only showed cases where the BH's net charge was chosen to be negative and of the order of $Q\approx 10^{-21}$, which results in $eQ \approx 1$ for electrons.
    
    For our stationary analytical model we needed to neglect a number of physical effects in the accretion process, which we detailed in the description of the model in Sec. \ref{setup}. In particular we neglected all particle interactions and the accretion disc's mass and charge. Where we consider charged dust we assume it to form a plasma consisting of noninteracting electrons and protons, which serve as test particles.
    For a stellar BH with a net charge of $10^{-18}$ -- $10^{-21}$ this restricts the plasma density to values smaller than $10$--$0.01\,\rm{cm}^{-3}$.
    %For a stellar BH with a net charge of $10^{-18}$ -- $10^{-21}$ this restricts the plasma density to values smaller than $10$--$0.01\,\rm{cm}^{-3}$. If $\hat r_0=\frac{r_0}{M}$ is fixed, the density has to be even more dilute for more massive BHs. This comes from the fact, that the total charge of the accreted particles, scales with $M^3$, while the total charge $Q=M\,\hat Q$ of the BH only scales with $M$.
    
   We analyzed the influence of the different parameters in our model on the accretion process and on the outer and inner edges of the forming accretion disc. Besides the density field, which we calculated numerically, all quantities -- namely the streamlines, the velocity field, the outer edge $r_{\rm D}$ and the ISCO -- were derived analytically. Four different values for the ISCO can be found for charged particles and a given BH spin and charge. These are connected to the four different combinations of same or opposite charge of BH and particles and direct or retrogating orbits. The ISCO is used to determine the inner edge of the accretion disc.
   
   We found that the spacetime parameters $a$ and $Q$ corresponding to the angular momentum and the charge of the BH, respectively, have a rather small effect on the accretion process and the edges of the accretion disc. However, we recovered the frame-dragging effect due to the angular momentum $a$ which was already discussed in \citep{acck} within our model. We showed that the product of BH and particle charge $eQ$, as well as the initial conditions for the $r$ and $\phi$ motion have a considerably stronger influence on the accretion process and the edges of the accretion disc than the spacetime parameters.
    
    When considering plasma contributing to or forming an accretion disc, we discussed four different cases which may occur within our model. In the first case all accreted particles will have to spiral into the BH. In the second and fourth case all particles of one type have to spiral into the BH, while a majority of the other particle type can contribute to the accretion disc. In this case the forming accretion disc might slowly develop a charge with the same or opposite sign as that of the BH until the arising electromagnetic field of the disc can not be neglected anymore and the model's description breaks down. However the case where an accretion disc with a charge opposite to that of the BH is developed seems rather unlikely due to expected interaction processes between the charged particles on the accretion disc, which are neglected in our model. In the third case a majority of both particles will contribute to or form an accretion disc. An inhomogeneous distribution of the charge of the disc is the result, where particles with a charge opposite to that of the BH are located in an inner area close to the BH, whereas particles with a charge with the same sign as that of the BH are located in an outer area farther away from the BH. This effect might be weakened or washed out due to the particle interactions within the accretion disc. Again, the model's description will break down, as soon as the arising electromagnetic fields from the charged areas are not negligible anymore.
    
\appendix
\section{Derivation of the solution for the particle motion}
\label{app1}
 In this appendix we present the derivation for the radial and longitudinal equation of motion for charged test particles in Kerr--Newman spacetime by using Jacobi elliptic functions. A comprehensive discussion of the solutions of 
  the Kerr--Newman equations of motions was done by Hackmann and Xu (2013), where they used Weierstrass elliptic functions \cite{hackmann}. More information on Jacobian elliptic functions can be found in \citet{abram} and \citet{byrd}.

  Elliptic integrals can take the form 
  \begin{equation} u(\phi)=\int_{\phi_0}^\phi\frac{dz}{\sqrt{P(z)}},\end{equation}
  where $P(z)$ is a polynomial of order three or four. The inverse function $\phi(u)$ of an elliptic integral is called elliptic function and it satisfies the differential equation
  \begin{equation}
  \left(\frac{d\phi}{du}\right)^2=P(\phi).
  \label{dglgl}
  \end{equation}
  This property is used to solve the differential equations \eqref{dqth} and \eqref{dqr} in terms of elliptic functions.
  
  We will now introduce elliptic integrals $F(\varphi,k)$ of the first kind, whose inverse functions are the Jacobi elliptic functions. They can take different forms, depending on which substitution is made for $\varphi$,
  \begin{align}
    F(\varphi,k)=&\int_0^\varphi\frac{d\vartheta}{\sqrt{1-k^2\sin^2(\vartheta)}}\nonumber\\
                =&\int_0^y\frac{dt}{\sqrt{(1-t^2)(1-k^2t^2)}} 
    \label{lnf}
  \end{align}
  for $y=\sin\varphi$. The parameter $k\in \mathbb{C}$ is called the modulus of the elliptic integral. The second integral in \eqref{lnf} with the polynomial under the square root of the form $P(t)=\sqrt{(1-t^2)(1-k^2t^2)}$ is called Legendre normal form, which only contains terms with even exponents.

  The Jacobian elliptic functions used in this paper are now defined as 
  \begin{align}
    \text{sn}(F,k) & =\sin\varphi = y\,,\\
    \text{cn}(F,k) & =\cos\varphi = \sqrt{1-y^2}\,.
  \end{align}
  Elliptic functions are doubly periodic and meromorphic, and the periods of $\text{sn}$ are given by $4K(k)$ and $4iK'(k)$, with
  \begin{align}
   K(k)=&\int_0^{\pi/2}\frac{d\vartheta}{\sqrt{1-k^2\sin^2(\vartheta)}}
  \end{align}
  and $K'(k)=K(k')$, $(k')^2 = 1-k^2$. 
  
  To derive the solutions for the $r$ and $\theta$ motion, we first derive two equations for the Mino time in terms of elliptic integrals from Eqs. \eqref{dqth} and \eqref{dqr},
   \begin{align}
     \label{gamma1}
     \lambda(r) =& \int_{r_0}^r\frac{dr'}{\sqrt{\textbf{R}(r')}},\\
     \lambda(\theta)=&\int_{\theta_0}^\theta\frac{d\theta'}{\sqrt{\Theta\left(\theta'\right)}}.
     \label{gamma2}
   \end{align}
   By introducing 
   \begin{align}
   \Phi(r)= \int_{r_a}^r \frac{dr'}{\sqrt{\textbf{R}(r')}}~,\\
   \Psi(\theta)=\int_{\theta_a}^\theta \frac{d\theta'}{\sqrt{\Theta\left(\theta'\right)}}~,
   \end{align}
   where $\textbf{R}(r_a)=0$ and $\Theta\left(\theta_a\right)=0$, we can rewrite Eqs. \eqref{gamma1} and \eqref{gamma2} as $\lambda(r)=\Phi(r)-\Phi(r_0)$ and $\lambda(\theta)=\Psi(\theta)-\Psi(\theta_0)$. To find the solutions of the $r$ and $\theta$ motion in terms of Jacobian elliptic functions we convert the polynomials $\textbf{R}(r)$ and $\Theta\left(\theta\right)$ to the Legendre normal form. This can be accomplished with substitutions of the form
   \begin{align}
   z & =\frac{A_1+A_2x^2}{A_3+A_4x^2}\, \text{ or }  \label{subs1}\\
   z & =\frac{B_1+B_2x}{B_3+B_4x}\,,\label{subs2}
   \end{align}
   with $z=r$ or $z=\theta$, respectively, and the constants $A_{1..4}$, $B_{1..4}$ have to be chosen properly.
   
   For the radial equation of motion the substitution \eqref{subs1} with $r=\frac{r_dx^2-n r_a}{x^2-n}$ is appropriate, where $r_{a..d}$ are the roots of $\textbf{R}(r)$. Now $k$ and $n$ have to be chosen such that the interval $r_1<r<r_2$, where the $r$-motion takes place, lies between $x=0$ and $x=1$. As a result we get, using the labeling of the roots mentioned in Sec. \ref{streamlines}, 
   \begin{align}
     \Phi(r)  &= \int_{r_a}^r \frac{dr}{\sqrt{\textbf{R}(r)}} \nonumber\\
      &= \frac{2}{\sqrt{(E^2-1)(r_a-r_c)(r_b-r_d)}} \nonumber\\ 
     &\quad\times \int_{0}^x \frac{dx}{\sqrt{(1-x^2)(1-k_r^2x^2)}}~,
   \end{align}
   where
   \begin{align}
     n & =\frac{r_d-r_b}{r_a-r_b}\,,\\
     k_r^2 & =\frac{(r_c-r_d)(r_a-r_b)}{(r_a-r_c)(r_b-r_d)}.
   \end{align}
   Now the solution for the radial motion can be written down,
    \begin{align}
	r(\lambda)=& \frac{r_a(r_d-r_b)+r_d(r_b-r_a)\,\text{sn}\left(\xi, k_r\right)^2}{r_d-r_b+(r_b-r_a)\,
	\text{sn}\left(\xi, k_r\right)^2},\\
	=& \frac{r_b(r_d-r_a)-r_d(r_b-r_a)\,\text{cn}\left(\xi, k_r\right)^2}{r_d-r_a-(r_b-r_a)\,
	\text{cn}\left(\xi, k_r\right)^2},
    \end{align}
    with 
    \begin{align}
    \xi=\frac{1}{2}\sqrt{(E^2-1)(r_a-r_c)(r_d-r_b)}\left[\Phi(r_0)-\lambda\right].
    \end{align}
    In case that all roots of $\textbf{R}(r)$ are real, the value of $\Phi(r)$ is always real and no complex numbers occur during the calculation of $r(\lambda)$. However in case of two or four complex roots the integrand of $\Phi(r)$ becomes complex and the calculation of $r(\lambda)$ has to be done in the complex plane. This is no problem in principle but can be avoided by using the substitution \eqref{subs2} and a wise choice of $B_{1..4}$. The exact substitution for these cases can be found in \cite{byrd} and will not be given here.
  
    For the $\theta$ motion we use the substitution $x=\cos{\theta}$ to get a polynomial of order four in $\Psi(\theta)$,
    \begin{align}
      \Psi(x)=&\int_{x_a}^x\frac{dx'}{\sqrt{\Theta\left(x'\right)}}\,,
    \end{align}
    for
    \begin{align}
    \Theta\left(x\right) & = a^2(1-E^2)x^4-(C+a^2(1-E^2)+ l^2)x^2 +C.
    \end{align}
    This can in general be solved by the same procedure used for solving the radial equation. To reduce the equation above to Legendre normal form, its roots have to be shifted to $1$ and $1/k^2$. Substituting $\tilde{x}=x/x_a$ then leads to 
    \begin{align}
      \Psi(\tilde{x})=-\frac{\cos\theta_a}{\sqrt{C}} \int_1^{\tilde{x}} \frac{dx'}{\sqrt{(1-{x'}^2)(1-\tilde{k}_\theta^2{x'}^2)}}\,,
    \end{align}
    with
    \begin{align}
     \tilde{k}_\theta^2=-\frac{a^2(E^2-1)}{C}x_a^4\,.
    \end{align}
    By rewriting $\tilde{k}_\theta$ as $\tilde{k}_\theta^2=\frac{{k_\theta}^2}{1-{k_\theta}^2}$ we find 
    \begin{align}
      \Psi(\tilde{x})&=-\frac{\cos\theta_a}{\sqrt{C+a^2(E^2-1)\cos^4\theta_a}}\nonumber\\
                     &\quad\times \int_1^{\tilde{x}} \frac{d x'}{\sqrt{(1-{x'}^2)(k_\theta'+k_\theta^2{x'}^2)}}\nonumber\\
	             &=\frac{\cos\theta_a\,\text{cn}^{-1}\left(\frac{\cos\theta}{\cos\theta_a},k_{\theta}\right)}{\sqrt{C+(E^2-1)a^2\cos^4\theta_a}}.
    \end{align}
    Finally the solution for the $\theta$ motion can be written down,
    \begin{align}
      &\cos(\theta(\lambda))=\nonumber\\ 
      &\cos\theta_a \text{cn}\left(\frac{\sqrt{C+a^2(E^2-1)\cos^4\theta_a}}{\cos\theta_a}\left(\Psi(\theta_0)-\lambda\right),k_\theta\right). \label{thetasol}
    \end{align}
    In the calculation of $\Psi(\theta)$ or $\cos(\theta(\lambda))$ complex numbers arise, if $k_\theta$ is imaginary or bigger than one. This can be avoided by choosing another Jacobian elliptic function to solve the equation and, by doing so, introducing a new $k_\theta$, which is then real and smaller than one. Again, we will not discuss this alternative descriptions of $\Psi(\theta)$ and $\cos(\theta(\lambda))$, but refer to \cite{byrd}.
%%%%%%%%%%%%%%%%%%%%%%%%%%%%%%%%%%%%%%%%%%%%%%%%%%
\section{ISCO in Kerr--Newman spacetime}
\label{app2}
  The innermost stable orbit (for $P=0$) is located where the effective potential of the $r$-motion $V_{\rm eff} (r) = \textbf{R}(r)$, see Eq. \eqref{dqr}, and its first and second derivative with respect to $r$ vanish. By further demanding that the ISCO is located in the equatorial plane, the $\theta$-motion $\frac{d \theta}{d \lambda}=\Theta(\theta)$ has to vanish at $\theta=\pi/2$. Therefor one receives an expression for the ISCO on the equatorial plane by solving the following 
  set of equations for coordinate $r$ and the three constants of motion $E,L$ and $K$:
  \begin{align}
    \textbf{R}(r) =&0 ,\nonumber\\ 
    \textbf{R}'(r)=&0 ,\nonumber\\
    \textbf{R}''(r)=&0 ,\nonumber\\ 
    \Theta(\pi/2) =&0.
    \label{soe}
  \end{align}
  From the last equation in \eqref{soe} the relation 
  \begin{equation}
    K=(aE-l)^2
  \end{equation}
  results. With this relation and the first two equations in \eqref{soe} a polynomial of order four can be deduced for $\sqrt{K}$ of the form
  \begin{align}
    f_1(u,\sqrt{K})&=A \sqrt{K}^4 +B \sqrt{K}^3+ C \sqrt{K}^2 \nonumber\\
                   &\quad + D \sqrt{K} +E=0,
  \end{align}
  where $u=1/r $, and
  \begin{align}
    A&=  \left(4 Q^4 + 4 Q^2 a^2\right) u^6+\left(-12 Q^2-4 a^2\right) u^5\nonumber\\
     & \quad+  \left(4 Q^2+9\right) u^4-6 u^3+u^2, \nonumber\\
    B&=   4 a eQ u^3 \left(Q^2 u^2 + a^2 u^2-2 u + 1\right),\nonumber\\
    C&=   \left(4 Q^4 + 4 Q^2 a^2 - Q^2 (eQ)^2-a^2 (eQ)^2\right) u^4\nonumber\\ 
     & \quad+   \left( -10 Q^2 - 2 a^2 + 2 (eQ)^2\right) u^3 \nonumber\\ 
     & \quad+   \left(2 Q^2 - 2 a^2-(eQ)^2 + 6\right) u^2-2 u,\nonumber\\
    D&=   B\, u^2 ,\nonumber\\
    E&=   \left(Q^4 + 2 Q^2 a^2-Q^2 (eQ)^2 + a^4 - a^2 (eQ)^2\right) u^2 \nonumber\\
     & \quad+  \left(-2 Q^2 - 2 a^2 + 2 (eQ)^2\right) u-(eQ)^2 + 1.    
  \end{align}
  Another equation $f_2(u,\sqrt{K})=0$ can be deduced from the second and third Eq. in \eqref{soe}. If we consider a very small charge of the BH, as explained in 
  Sec. \ref{setup}, and set $Q=0$ but $eQ\neq 0$, $f_2(u,\sqrt{K})$ reduces to
  \begin{align}
    f_2(u,\sqrt{K})&=  6\left( a \sqrt{K}^3-a^2 eQ K\right) u^2 \nonumber\\
                   & \quad+ 6\left(a (eQ)^2 \sqrt{K} - eQ K \right) u \nonumber \\
                   & \quad+  eQ\left(a^2 +K \right ) - (eQ)^3 - 2 a \sqrt{K}=0.
  \end{align}
  Now $f_1(u,\sqrt{K})$ (for $Q=0$, $eQ\neq 0$) and $f_2(u,\sqrt{K})$ can be solved numerically for $u=1/r$ and $K$ and lead to four different solutions for every given set of parameters $(a,eQ)$. 

\begin{acknowledgments}
    The authors thank Emilio Tejeda and Volker Perlick for insightful discussions. Support from the DFG through the Research Training Group 1620 'Models of Gravity' and the Collaborative Research Center 1128 'Relativistic Geodesy and Gravimetry with Quantum Sensors (geo-Q)' is gratefully acknowledged.
\end{acknowledgments}

\bibliography{sample.bib}

\end{document}